\documentclass[a4paper,12pt]{article}

\usepackage{amsmath}
\usepackage{amssymb}
\usepackage{graphicx}
\usepackage{hyperref}
\usepackage{cite}
\usepackage{color}

\makeatletter
\@addtoreset{equation}{section}
\renewcommand{\theequation}{\thesection.\@arabic\c@equation}
\makeatother

\newcommand{\rshell}{r^{\tiny sh}}
\newcommand{\Cshell}{C_{\tiny sh}}

\definecolor{green}{rgb}{0,0.5,0}

\begin{document}

\begin{titlepage}

\baselineskip 24pt   
\vglue 10mm   

\begin{center}
{\Large\bf
Static Black Hole and Vacuum Energy: \\
Thin Shell and Incompressible Fluid
\\
}

\vspace{8mm}   

\baselineskip 18pt   

\renewcommand{\thefootnote}{\fnsymbol{footnote}}

Pei-Ming Ho
\footnote[2]{e-mail address: pmho@phys.ntu.edu.tw},
Yoshinori Matsuo%
\footnote[3]{e-mail address: matsuo@phys.ntu.edu.tw}

\renewcommand{\thefootnote}{\arabic{footnote}}
 
\vspace{5mm}   

{\it  
Department of Physics and Center for Theoretical Sciences, \\
National Taiwan University, Taipei 106, Taiwan,
R.O.C. 
}
  
\vspace{10mm}   

\end{center}

\begin{abstract}

With the back reaction of the vacuum energy-momentum tensor consistently taken into account,
we study static spherically symmetric black-hole-like solutions
to the semi-classical Einstein equation.
The vacuum energy is assumed to be given by that of
2-dimensional massless scalar fields, 
as a widely used model in the literature for black holes.
The solutions have no horizon.
Instead, 
there is a local minimum in the radius.
We consider thin shells 
as well as incompressible fluid
as the matter content of the black-hole-like geometry.
The geometry has several interesting features
due to the back reaction of vacuum energy.
In particular, Buchdahl's inequality can be violated without divergence in pressure, 
even if the surface is below the Schwarzschild radius. 
At the same time, the surface of the star can not be far below the Schwarzschild radius 
for a density not much higher
than the Planck scale, 
and the proper distance from its surface to the origin can be very short 
even for very large Schwarzschild radius. 
The results also imply that,
contrary to the folklore,
in principle the Boulware vacuum can be physical
for black holes. 

\end{abstract}

\end{titlepage}

\newpage

\baselineskip 18pt

\noindent\rule{\textwidth}{1pt}

\tableofcontents

\vskip 12pt

\noindent\rule{\textwidth}{1pt}



\section{Introduction}

What is a black hole?
The notion of event horizon used to play an important role in
our understanding of black holes.
Nowadays,
the event horizon is considered by many as 
an inappropriate concept for physical black holes,
as its experimental verification takes an infinitely long time.
In fact,
even the necessity of apparent horizon for black holes has been questioned
\cite{Gerlach:1976ji,FuzzBall,FuzzBall2,Barcelo:2007yk,Vachaspati:2006ki,Krueger:2008nq,Kawai:2013mda,Kawai:2014afa,Ho:2015fja,Kawai:2015uya,Ho:2015vga,Ho:2016acf,Kawai:2017txu,Fayos:2011zza,Mersini-Houghton,Saini:2015dea,Baccetti}.
In this paper,
we will refer to back-hole-like objects simply as black holes.

Recently,
by taking into account self-consistently the back reaction from Hawking radiation
through the semi-classical Einstein equation,
it was shown
\cite{Kawai:2013mda,Kawai:2014afa,Ho:2015fja,Kawai:2015uya,Ho:2015vga,Ho:2016acf,Kawai:2017txu} 
that,
if the vacuum energy near the Schwarzschild radius
is dominated by Hawking radiation,
neither the event nor the apparent horizon forms during a gravitational collapse.
Later,
it was also shown \cite{Ho:2017joh} that
even for static black holes, 
for which there is no Hawking radiation,
the vacuum energy-momentum tensor is capable of
removing the horizons.%
\footnote{
For static, spherically symmetric configurations,
the event horizon, apparent horizon and Killing horizon coincide.
}
Different models of quantum vacuum energy leads to different near-horizon geometries.
For some of the models of vacuum energy,
there is no horizon around the Schwarzschild radius
and the near-horizon geometry is replaced by
a (traversable) wormhole-like structure, 
that is,
a local minimum of the radius $r$.%
\footnote{
For spherically symmetric space-time,
we define the radius $r$ for a symmetric sphere
such that its area is $4\pi r^2$.
}

In Ref.\cite{Ho:2017joh},
only the vacuum solution is considered.
In a more realistic model, 
the wormhole-like structure does not continue to another open space,
and the vacuum solution only applies to the exterior of a star of finite radius.
The geometry of the vacuum solution should be
put in junction with that for a matter distribution in the star,
where the radius continues to zero at the center of the star.

In this paper,
we study the static geometry inside the neck 
(local minimum of the radius) of the wormhole-like structure 
as well as the internal space of the star,
assuming spherical symmetry. 
Using non-perturbative methods,
we uncover novel features of the black-hole geometry 
that cannot be captured in the perturbative approach.

We first consider a star composed of a spherical thin shell 
whose energy distribution is proportional to a delta function of the radial coordinate.
Next, we consider a star which consists of incompressible fluid,
as a simple example of continuous distribution of matter. 
Since it is difficult to solve the semi-classical Einstein equation exactly, 
we analyze the solution in different regions with approximate
analytic as well as numeric solutions.

The solutions for a star composed of incompressible fluid
depend on three physical parameters.
They can be chosen to be
the Schwarzschild radius (or the total mass),
the energy density of the fluid
and the size (or the radius of the outer surface) of the star.
We provide a numerical method that allows us to explore
the relation among these three parameters
and study the distribution of mass and pressure in the star.

These solutions demonstrate interesting differences from their classical counterparts
due to the back reaction of vacuum energy.
For instance,
Buchdahl's inequality ($r > 9a/8$) can be violated
without divergence in pressure.
Furthermore,
the surface of the star can never be far below the ``neck''
--- the local minimum of the radius $r$,
which is close to the Schwarzschild radius, 
as long as the density of the incompressible fluid is not much larger than the Planck scale. 
The reason for this conclusion is a peculiar feature of the vacuum solution under the neck,
and is independent of the matter content of the star.
The result that the solutions have no singularity nor horizon 
implies that an arbitrarily heavy star can have 
a physical state in the Boulware vacuum as long as it has stationary states, 
in contrary to the folklore that the Boulware vacuum is unphysical for black holes. 
The folklore says that the Boulware vacuum is unphysical 
if the radius of the star is smaller than the Schwarzschild radius 
since the energy-momentum tensor diverges at the Schwarzschild radius. 
However, it is physically sensible to consider the Boulware vacuum for any macroscopic radius 
as is shown in \cite{Ho:2017joh} by 
nonperturbative analysis of the semi-classical Einstein equation. 
On the other hand, even if we only pay attention to the case when the radius of the star 
is larger than the Schwarzschild radius, we have obtained interesting nonperturbative results.
For example, the surface is always outside the Schwarzschild radius 
as long as the density is much lower than the Planck scale. 

Our work also contributes new inputs to understand the information loss paradox.
While string theory provides strong evidence that the information is not lost,
the assumption that the information of the collapsing matter is encoded in
Hawking radiation leads to the conclusion that the vicinity of the horizon 
cannot be largely empty 
or 
free of order-1 corrections to the classical geometry.
Otherwise,
the low energy effective theory should be valid,
which implies that Hawking radiation cannot carry all the information.
(See \cite{Mathur:2009hf} and \cite{Marolf:2017jkr}
for more discussions about the information loss paradox.)
In Mathur's words \cite{Mathur:2009hf},
some ``niceness conditions'' must be violated to invalidate the low energy effective theory.
Examples of such proposals include the fuzzball \cite{FuzzBall} and the firewall \cite{Almheiri:2012rt}.
However, 
it is still unclear how in detail a gravitational collapse leads to configurations
for which low energy effective theories are invalid.
An important implication of our results is that the star has a finite 
but large (possibly Planck scale) energy density and pressure 
when the radius is not much larger than the Schwarzschild radius. 
This is an indication that a low energy effective theory is insufficient.
Our model is an exception 
to the conventional view that there is a large empty space
(with a negative energy for the vacuum state) around the Schwarzschild radius. 

This paper is organized as follows. 
In Section~\ref{sec:model}, 
we introduce a model for the vacuum energy that has been widely used in the literature
(see, e.g.\
\cite{Davies:1976ei,Christensen:1977jc,Parentani:1994ij,Brout:1995rd,Ayal:1997ab,Fabbri:2005nt,Barcelo:2007yk}),
to study the back reaction of vacuum energy for a black hole.
According to the semi-classical Einstein equation,
the geometry around the Schwarzschild radius is modified by the vacuum energy,
and relevant results of Ref.\cite{Ho:2017joh} are briefly reviewed in Section~\ref{sec:exterior}.
For the cases when the star is further below the neck,
we give an approximate analytic solution to the vacuum geometry below the neck and above the star.
The new results are give in Section~\ref{sec:behindtheneck}.
Interestingly,
the peculiar geometry there does not allow the star to be far below the neck.
In Section~\ref{sec:shell}, 
we consider a star composed of a thin shell.
In Section~\ref{sec:fluid},
we introduce the model of a star composed of an incompressible fluid. 
We prove analytically that the geometry does not have singularity
except possibly at $r = 0$.
Solutions violating Buchdahl's inequality are regularized by 
the back reaction of vacuum energy.
In Section~\ref{sec:numerical}, 
we show the results of numerical calculations. 
Section~\ref{sec:conclusion} is devoted to conclusion and discussions. 
While the model of vacuum energy under consideration may or may not
be a good approximation to the real world,
this work provides a concrete model for the possibility 
of an interesting scenario in which
the back reaction of vacuum energy plays an important role
to the black-hole geometry.


\section{The model}\label{sec:model}

In this paper,
we focus on the 4-dimensional semi-classical Einstein equation,%
\footnote{
The validity of the semi-classical Einstein equation is one of our major assumptions.
Some proposed that the back reaction problem for black holes 
should be treated in terms of the stochastic gravity \cite{StochasticGravity}.
}
\begin{equation}
 G_{\mu\nu} = \kappa \langle T_{\mu\nu} \rangle \ , 
\label{Einstein}
\end{equation}
where $G_{\mu\nu}$ is the classical Einstein tensor, 
while the expectation value of the energy-momentum tensor 
$\langle T_{\mu\nu}\rangle$ contains the quantum effect. 
We assume that the energy-momentum tensor can be separated as 
\begin{equation}
 \langle T_{\mu\nu} \rangle = T^m_{\mu\nu} + T^\Omega_{\mu\nu} \ ,
\end{equation}
where $T_{\mu\nu}^m$ is the classical energy-momentum tensor of matter 
and $T_{\mu\nu}^{\Omega}$ represents the quantum effect.
Typically, the latter is calculated as the vacuum expectation value
of the energy-momentum operator of certain quantum fields.

Here, we consider only static spherically symmetric configurations 
and so a generic metric can be given in the form
\begin{equation}
 ds^2 = -C(r) dt^2 + \frac{C(r)}{F^2(r)} dr^2 + r^2 d \Omega^2 \ .
\label{metric}
\end{equation}
Defining the tortoise coordinate $r_*$ by 
\begin{equation}
dr_* = \frac{dr}{F(r)},
\label{tortoise}
\end{equation}
we can express the metric as 
\begin{equation}
 ds^2 = - C(v-u) du\,dv + r^2(v-u) d \Omega^2
\end{equation}
in terms of the null coordinates defined by
\begin{align}
 u &= t - r_* \ , & 
 v &= t + r_*  \ , 
\end{align}

Following Refs.\cite{Davies:1976ei,Christensen:1977jc,Parentani:1994ij,Brout:1995rd,Ayal:1997ab,Fabbri:2005nt,Barcelo:2007yk},
we consider the model for the vacuum energy-momentum tensor $T^\Omega_{\mu\nu}$ 
defined by
\begin{equation}
 T^\Omega_{\mu\nu} = \frac{1}{r^2} T^{(2D)}_{\mu\nu} \ , \label{VEM}
\end{equation}
where $T^{(2D)}_{\mu\nu}$ is the vacuum expectation value of
the energy-momentum tensor operator of 
$N$ 2-dimensional massless scalar fields \cite{Davies:1976hi}
obtained through a spherical reduction of the 4-dimensional space-time.
$T^{(2D)}_{\mu\nu}$ is completely determined by the Weyl anomaly and conservation law 
up to the initial (or boundary) conditions. 
In terms of the null coordinates, 
the energy-momentum tensor of the 2-dimensional scalars is given by
\begin{align}
 T_{uu}^{(2D)} &= - \frac{1}{12\pi} C^{1/2} \partial_u^2 C^{-1/2} + \widehat T_{uu}(u) \ , 
\label{VEMuu}
\\
 T_{vv}^{(2D)} &= - \frac{1}{12\pi} C^{1/2} \partial_v^2 C^{-1/2} + \widehat T_{vv}(v) \ ,
\label{VEMvv}
\\
 T_{uv}^{(2D)} &= 
 \frac{1}{24\pi C^2} \left(\partial_u C \partial_v C - C \partial_u \partial_v C \right) \ .
\label{VEMuv}
\end{align}
The single-variable functions $\widehat T_{uu}(u)$ and $\widehat T_{vv}(v)$ are
the integration ``constants'' arising from solving the conservation law.
They should be fixed by the initial (boundary) conditions. 
Here, we focus on the static configurations 
without any incoming or outgoing energy flow at the spatial infinity ($r\to\infty$). 
Choosing the gauge in which $C\to 1$ in the limit $r\to\infty$,
we have
\begin{equation}
 \widehat T_{uu}(u) = \widehat T_{vv}(v) = 0 \ . 
\label{ZeroT}
\end{equation}
The quantum state corresponding to this boundary condition 
is called the Boulware vacuum \cite{Boulware}.
It is suitable for describing static configurations.

As the lowest order approximation of a perturbation theory,
the energy-momentum tensor $T^\Omega_{\mu\nu}$ of the Boulware vacuum
was calculated for the Schwarzschild background \cite{Davies:1976ei},
and found to diverge at the horizon.
Hence,
conventionally,
the Boulware vacuum is used only for static stars
whose radii are larger than the Schwarzschild radius.
However,
it was recently shown \cite{Ho:2017joh} that
the perturbation theory breaks down at the horizon,
and nonperturbatively $T^\Omega_{\mu\nu}$ is non-singular
within a certain range around the Schwarzschild radius.

In the case of a static star of radius $R$ much larger than its Schwarzschild radius,
the conventional wisdom is that the Boulware vacuum is the appropriate quantum state
for a quantum field in this background.
Now imagine that we shrink the star adiabatically,
with every particle in the star moving extremely slowly towards the origin.
At any given time,
the geometry outside the star can be arbitrarily well approximated
by a static solution of the semi-classical Einstein equation,
assuming that the time scale of the change in radius can be arbitrarily long.
The assumption that the time scale of changes can be arbitrarily long
would break down if there is a horizon.
However,
as we have proven analytically in Ref.\cite{Ho:2017joh},
the geometry has no horizon due to the back reaction of the vacuum energy defined above.
It is therefore theoretically possible that
the Boulware vacuum can be applied to a star hidden below the Schwarzschild radius.
The goal of this paper is to understand the geometry of a static star
in the Boulware vacuum in more detail,
even when the star is submerged below the surface at the Schwarzschild radius.


\section{Geometry of vacuum outside the star}\label{sec:exterior}

In Ref.\cite{Fabbri:2005nt,Ho:2017joh},
it was shown that there is no horizon for a static star with spherical symmetry
in the Boulware vacuum due to the back reaction of the vacuum energy-momentum tensor
for the model described in the previous section. 
While the geometry outside the Schwarzschild radius remains almost identical 
to the Schwarzschild solution,
the horizon is deformed to a (traversable) wormhole-like geometry.%
\footnote{%
Wormhole geometries were proposed to be candidates of 
semi-classical black-hole geometries earlier in Refs.~\cite{Solodukhin:2004rv}.
}
More specifically, 
there is a turning point where the radius $r$ is at a local minimum.
We will refer to this point (a 2-sphere) of local minimum in $r$ as the ``throat'' or ``neck''
of the wormhole-like structure,
while it is also called a ``turning point'' or a ``bounce'' in the literature
\cite{Fabbri:2005nt}.
The radius $r = a$ of the neck will be called the ``quantum Schwarzschild radius''.
At large distance,
the geometry approaches to a Schwarzschild solution
with a certain Schwarzschild radius $r = a_0$,
which will be referred to as the ``classical Schwarzschild radius''.
Sometimes we will just say ``Schwarzschild radius''
if it does not matter which one we are referring to,
as the difference between the values of $a$ and $a_0$ is extremely small
for a large $a$. 

It should be noted that the total mass of the system is not related to 
the quantum Schwarzschild radius but to the classical Schwarzschild radius. 
Since the quantum Schwarzschild radius is defined at the neck of the wormhole-like structure, 
it does not include the effects of the vacuum energy outside the neck, 
while the classical Schwarzschild radius is related to the total mass of the system 
since it is defined by the asymptotic structure of the spacetime.  

Let us review the exterior geometry of a static star in vacuum.
The classical energy-momentum tensor for matter is zero outside the star:
\begin{equation}
 T_{\mu\nu}^m = 0 \ . 
\end{equation}
For static, spherically symmetric configurations, 
the semi-classical Einstein equation
(with the vacuum energy-momentum tensor given in the previous section)
gives the following differential equations for $C(r)$ and $F(r)$ in the metric \eqref{metric}:
\begin{align}
F C' - F' C
- \frac{\alpha}{2} \frac{1}{r} (F' C' + F C'')
+ \frac{3\alpha}{4} \frac{1}{C r} F {C'}^2 = 0 \ ,
\label{CF-1}
\\
\frac{C^2}{F r} - \frac{FC}{r} - F' C
- \frac{\alpha}{2} \frac{1}{r} (F' C' + F C'')
+ \frac{\alpha}{2} \frac{1}{C r} F {C'}^2
= 0 \ ,
\label{CF-2}
\end{align}
where $C$ and $F$ depend only on $r$ 
and a prime indicates the derivative with respect to $r$. 
The parameter $\alpha$ is defined by 
\begin{equation}
 \alpha = \frac{\kappa N}{24\pi} \ , 
\end{equation}
where $N$ is the number of massless scalar fields. 

In the semi-classical Einstein equations,
$\sqrt{\alpha}$ characterizes the length scale of the quantum correction.
For $N$ of order $1$,
$\sqrt{\alpha}$ is of the order of Planck length $\ell_p$.
For a very large $N$,
we can have 
$\sqrt{\alpha} \gg \ell_p$,
so that the quantum effect of the matter fields becomes important at a sub-Planckian scale
when the effect of quantum gravity is still suppressed.

Eqs.\eqref{CF-1} and \eqref{CF-2} are equivalent to a single differential equation \cite{Ho:2017joh}:
\begin{equation}
 2 r \rho'(r) + (2r^2+\alpha)\rho^{\prime\,2}(r) + \alpha r \rho^{\prime\,3}(r) 
 + (r^2 - \alpha) \rho''(r) = 0 \ , \label{eq-rho}
\end{equation}
with $C$ and $F$ given in terms of $\rho$ as 
\begin{align}
 C(r) 
 &= 
 e^{2\rho(r)} \ , 
\\
 F(r) 
 &= 
 \frac{e^{\rho(r)}}{\sqrt{1 + 2 r \rho'(r) + \alpha \rho^{\prime\,2}(r)}} \ . 
\end{align}

Eq.\eqref{eq-rho} is a second order differential equation.
Assuming asymptotic Minkowski space,
the solution space has 2 parameters.
One of the two parameters is the mass parameter 
(or the Schwarzschild radius)
of the approximate Schwarzschild solution at distance.
The other is just a scaling parameter corresponding to
a constant scaling of the coordinates $(t, r_*)$.
In other words,
given the Schwarzschild radius of the asymptotically Schwarzschild solution,
there is a unique spherically symmetric solution to the Einstein equation 
for the vacuum energy \eqref{VEMuu} -- \eqref{VEMuv}
with zero flux \eqref{ZeroT} at spatial infinity.

\subsection{The neck}
\label{sec:theneck}

Naively,
one expects that the Einstein equation \eqref{eq-rho} can be solved 
as a perturbative expansion in powers of $\kappa$, or equivalently $\alpha$. 
However, this perturbative expansion is not valid around the Schwarzschild radius. 
If we expand $\rho$ in terms of $\alpha$ as 
\begin{equation}
 \rho(r) = \rho_0(r) + \alpha \rho_1(r) + \cdots \ , 
\label{rho-expand}
\end{equation}
the 0-th and 1st order solutions $\rho_0$ and $\rho_1$ for eq.\eqref{eq-rho} are
\begin{align}
 \rho_0(r) 
 &= 
 \frac{1}{2} \log c_0 + \frac{1}{2} \log\left(1-\frac{a_0}{r}\right) \ , 
\label{pert-rho0}
\\
 \rho_1(r) 
 &= 
 - \frac{4 r^2 + a_0^2 + 4 a_0 r (2 c_1 r - 1)}{8 a_0 r^2 (r-a_0)} 
 - \frac{2 r - 3 a_0}{4 a_0^2 (r-a_0)} \log\left(1 - \frac{a_0}{r}\right) \ , 
\label{pert-rho1}
\end{align}
where $a_0$, $c_0$ and $c_1$ are integration constants. 
The constant $a_0$ is the classical Schwarzschild radius. 
The divergence in $\rho_0$ at $r=a_0$ implies that $C(r)$ goes to zero 
at $r=a_0$ in the classical limit.
But the divergence in $\rho_1$ implies that the quantum correction for $C(r)$ 
diverges at the horizon,
where the perturbative expansion breaks down. 
One should resort to non-perturbative approaches.

It was analytically proven \cite{Ho:2017joh}
that $C$ cannot go to zero at finite $r$,
due to the nonperturbative nature of eq.\eqref{eq-rho}. 
Let us briefly review the prove here.
While the expansion \eqref{rho-expand} is valid outside the Schwarzschild radius,
$\rho'$ increases indefinitely as $r$ decreases.
For sufficiently large $\rho'$,
the first and second term in eq.\eqref{eq-rho} can be neglected,
and the equation is then approximated by 
\begin{equation}
 \alpha r \rho^{\prime\,3}(r) + (r^2 - \alpha) \rho''(r) \simeq 0 \ . 
\end{equation}
Note that the first term would be absent if $\alpha = 0$,
i.e., if there were no quantum correction to the vacuum energy.

According to this equation,
the function $\rho'$ continues to increase as $r$ decreases
until $\rho'$ diverges at some point,
say, $r=a$. 
We call this radius the quantum Schwarzschild radius.
It is the radius of the ``neck'' of the wormhole-like structure \cite{Ho:2017joh}. 
The solution of $\rho$ can be expanded around $r=a$ as 
\begin{equation}
 \rho = \frac{1}{2}\log c_0 + \sqrt{k(r-a)} + \mathcal O(r-a) \ , 
\label{rho-sol}
\end{equation}
where the constant $k$ is given by 
\begin{equation}
 k \equiv \frac{2(a^2-\alpha)}{\alpha a} \simeq \frac{2a}{\alpha} \ , 
\end{equation}
and $c_0$ is an integration constant. 
We shall always assume that $a^2 \gg \alpha$.

The expression \eqref{rho-sol} is a good approximation for 
\begin{equation}
 r - a \ll \frac{\alpha}{a} \ , 
 \label{cond-r1}
\end{equation}
while the perturbative expansion \eqref{rho-expand}-\eqref{pert-rho1} 
is good outside the Schwarzschild radius for
\begin{equation}
 r - a_0 \gg \frac{\alpha}{a_0} \ . 
 \label{cond-r2}
\end{equation}
As these two approximation schemes are supposed to meet
around the points where $r-a \sim \mathcal O(\alpha/a)$, 
the quantum Schwarzschild radius $a$ and the classical Schwarzschild radius $a_0$ differ by
\begin{equation}
a - a_0 \sim {\cal O}\left(\frac{\alpha}{a}\right),
\end{equation}
and the order of magnitude of the constant $c_0$ can be roughly estimated as \cite{Ho:2017joh}
\begin{equation}
 c_0 \sim \mathcal O\left(\frac{\alpha}{a^2}\right) \ . 
\end{equation}

Around the quantum Schwarzschild radius $r=a$,
the metric is approximately given by \cite{Ho:2017joh}
\begin{equation}
 ds^2 \simeq - c_0 e^{2 \sqrt{k(r-a)}} dt^2 + \frac{\alpha k dr^2}{4 (r-a)} + r^2 d \Omega^2 \ . 
\label{metric-wormhole}
\end{equation}
The proper length of the throat region approximated by this metric
(for $0 \leq r - a \ll \frac{\alpha}{a}$)
is of the order of magnitude of
\begin{equation}
\Delta s < \int_a^{a + {\cal O}(\alpha/a)} dr \, \sqrt{\frac{a}{2(r-a)}}
< {\cal O}(\sqrt{\alpha}).
\label{near-neck-length}
\end{equation}

In terms of the tortoise coordinate $r_*$,
the metric
\eqref{metric-wormhole} becomes
\begin{align}
 ds^2 &\simeq - \left[c_0 e^{2\sqrt{k(r-a)}}  + {\cal O}((r_* - a_*)^2)\right] (dt^2 - dr_*^2) 
\notag\\&\quad 
 + \left[a^2 + \frac{2 a c_0}{\alpha k} \left(r_* - a_*\right)^2 + 
 {\cal O}((r_* - a_*)^3)\right] d\Omega^2 \ , 
\label{metric-near}
\end{align}
where $a_*$ is the value of the tortoise coordinate $r_*$ when $r=a$. 
Examining the coefficient of the term $d\Omega^2$,
we see that $r=a$ is the minimum of the radius $r$.
(The radius $r$ is defined such that
the area of a symmetric 2-sphere equals $4\pi r^2$.)
This geometry resembles a traversable static wormhole
whose ``neck'' or ``throat'' is a local minimum of $r$.
It is certainly not a genuine wormhole,
as the vacuum inside the neck terminates on the surface of the star,
instead of leading to another open spacetime.

Due to the back reaction of vacuum energy,
the horizon at the classical Schwarzschild radius $a_0$ is replaced by a wormhole-like geometry
with the neck at the quantum Schwarzschild radius $r = a$.
Notice that even though $\partial_v r = 0$ at $r = a$,
suggesting that the outgoing null vectors normal to the neck are non-expanding,
it is not an apparent horizon since it is not the boundary of a trapped region.
It is simply a local minimum of $r$.

The energy density around the neck of the wormhole is estimated as 
\begin{equation}
 - \langle T^0{}_0 \rangle 
 \simeq - \frac{1}{\kappa a^2} \left(1 - 6 \sqrt{k(r-a)} + \cdots \right) \ . 
\end{equation}
This is of the same order as the naive non-relativistic average mass density for a star of radius a; 
\begin{equation}
\frac{\mbox{mass}}{\mbox{volume}} \sim \frac{\frac{4\pi a}{\kappa}}{\frac{4\pi}{3}a^3}
\sim {\cal O}(a^{-2}\ell_p^{-2}) \ , 
\end{equation}
but it is much smaller than the mass density of the matter
in a solution to the semi-classical Einstein equation,
which is of order ${\cal O}(\kappa^{-1}\alpha^{-1})$.
(See eq.\eqref{mak} below.)
Its contribution to the total mass is negligible because of an additional redshift factor; 
\begin{equation}
 \Delta M 
 \sim a^2 \Delta s \sqrt{c_0} \, \langle T^0{}_0 \rangle 
 < \mathcal O(a^{-1} \ell_p^0) \ . 
\end{equation}

\subsection{Behind the neck}
\label{sec:behindtheneck}

The metric \eqref{metric-wormhole} (or \eqref{metric-near}) is valid 
in a small region around the neck
when the radius $r$ satisfies \eqref{cond-r1}.
In this subsection,
we assume that the surface of the star is further deeper down the neck,
where \eqref{cond-r1} is no longer satisfied,
and we study the geometry with back reaction from the vacuum energy.

As we move down the neck towards the star,
the radius $r$ increases,
and the magnitudes of $\rho'$ and $\rho''$ decrease 
according to eq.\eqref{rho-sol},
with $\rho''$ decreasing faster than $\rho'$,
until we reach the surface of the star.
When the condition \eqref{cond-r1} is no longer valid,
eq.\eqref{eq-rho} is dominated by the $\rho^{\prime\, 2}$-term and the $\rho^{\prime\, 3}$-term:
\begin{equation}
 2r^2 \rho^{\prime\,2}(r) + \alpha r \rho^{\prime\,3}(r) \simeq 0 \ , 
\end{equation}
and so we find the approximate solution
\begin{align}
\rho'(r) \simeq - \frac{2r}{\alpha}
\label{eq-linear-rhop}
\end{align}
for the vacuum space below the neck of the wormhole-like geometry when $r - a \gg \alpha/a$.%
\footnote{
Note that this condition $r - a \gg \alpha/a$ refers to the region below the neck, 
while eq.\eqref{cond-r2} refers to the region above the neck.
}

The behavior \eqref{eq-linear-rhop}
can be better understood as follows.
First, we note that eq.\eqref{eq-linear-rhop} is a small deviation to the exact solution
\begin{equation}
 \rho'(r) = - \frac{1}{\alpha}\left( r + \sqrt{r^2 - \alpha}\right) \ , 
\label{rho-exact}
\end{equation}
to eq.\eqref{eq-rho} at large $r$ (below the neck).
The function $F(r)$ is singular for this solution \eqref{rho-exact},
but we will consider small perturbations of this solution so that $F$ is regular.
While eq.\eqref{eq-rho} is a second order differential equation of $\rho$,
it is a first order differential equation of $\rho'$.
Hence we expect to find solutions with an integration constant 
as deviations of the special solution \eqref{rho-exact} in the form:
\begin{equation}
 \rho'(r) = - \frac{1}{\alpha}\left( r + \sqrt{r^2 - \alpha}\right) + \delta\rho'(r) \ . 
\end{equation}
Substituting this into eq.\eqref{eq-rho} and expanding to the linear order in $\delta\rho'(r)$, 
the correction term $\delta\rho'(r)$ is solved as ok
\begin{equation}
 \delta\rho'(r) = 
 - f_0 e^{-\frac{2r^2}{\alpha}} \left(2 r^2 - \frac{5 \alpha}{4} 
 + \mathcal O(\alpha^2)\right) + \mathcal O(f_0^2) \ , 
 \label{eq-delta-rhop}
\end{equation}
where $f_0$ is an integration constant. 

Note that the factor $e^{-2r^2/\alpha}$ is extremely small
so that the value of $f_0$ can be very large while keeping 
the magnitude of the correction $\delta\rho'(r)$ sufficiently small
for a valid perturbative expansion.
The condition for $\delta \rho'$ to be a small perturbation is
\begin{equation}
|f_0| \, r \, e^{-{2r^2}/{\alpha}} < \frac{1}{\alpha}.
\end{equation}
Since the left hand side quickly decreases with increasing $r$,
the condition only needs to be checked near the neck.
Let the approximation \eqref{eq-delta-rhop} be valid for $r \geq r_1$,
where $r_1 - a \sim {\cal O}(\alpha/a)$,
we assume that the parameter $f_0$ satisfies the condition
\begin{equation}
|f_0| \, r_1 \, e^{-{2 r_1^2}/{\alpha}} < \frac{1}{\alpha}.
\label{cond-f0}
\end{equation}

The deviation $\delta\rho'(r)$ goes to zero as $r$ increases,
as a class of solutions approaching to the same attractor solution \eqref{rho-exact}.
This explains the robust linear bahvior \eqref{eq-linear-rhop}) observed in $\rho'$ at large $r$.

Eq.\eqref{eq-delta-rhop} implies that
\begin{equation}
 F(r) \simeq - \frac{1}{f_0^{1/2} r} \left(1+ \frac{\alpha}{2 r^2} 
 + \mathcal O(\alpha^2)\right) + \mathcal O(f_0^{1/2}) \ ,
\end{equation}
and the metric in this region is approximately
\begin{equation}
 ds^2 \simeq r e^{- 2r^2/\alpha} 
 \left[- \tilde c_0 \left(1- \frac{\alpha}{8r^2}\right) dt^2 
 + f_0 \left(r^2 - \frac{9}{8}\alpha\right) dr^2 \right] 
 + r^2 d \Omega^2 + \mathcal O(\alpha^2) \ . 
\label{DeepInside}
\end{equation}
Assuming that this metric can be connected to eq.\eqref{metric-wormhole}
around $r-a \sim \mathcal O(\alpha/a)$, 
we estimate $\tilde c_0$ and $f_0$ as 
\begin{align}
 e^{-2 a^2/\alpha} \tilde c_0 &\sim \mathcal O(\alpha/a^3) \ , & 
 e^{-2 a^2/\alpha} f_0 &\sim \mathcal O(a^{-1}\alpha^{-1}) \ . 
\end{align}

The scalar curvature 
\begin{equation}
 R \simeq	 - \frac{8 \alpha}{f_0 r^7} e^{2 r^2/\alpha} \ 
 \label{divergence-R}
\end{equation}
diverges in the limit $r\to\infty$ under the neck.
This singularity at infinite $r$ is in fact within finite proper distance from the neck.
But this divergence is irrelevant in physical situations
as it is a divergence of the vacuum solution.
In the more realistic case,
there is a star with a surface of finite radius,
and the vacuum solution does not apply to the internal region 
behind the surface of the star.
On the other hand,
this singularity is analogous to the singularity at the origin of the Schwarzschild solution.
The asymptotic Schwarzschild spacetime with a positive classical Schwarzschild radius $a_0$
implies that there is positive energy under the neck.
The Einstein equation with vacuum energy does not allow a positive mass except at $r \to \infty$.
Hence the singularity \eqref{divergence-R} is a reflection of the needed positive mass,
just like the singularity at the origin in the Schwarzschild solution.

\begin{figure}
\vskip1cm
\begin{center}
\includegraphics[scale=0.5,bb=0 0 375 280]{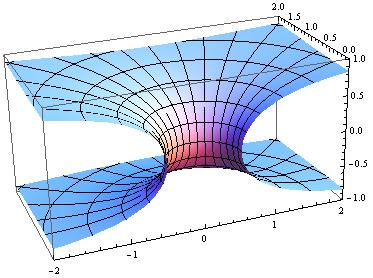}
\caption{\small 
The schematical cross section of a black hole around the neck,
assuming that the surface of the star is under the neck
(not shown).
The radius $r$ has a local minimum at the quantum Schwarzschild radius $a$,
and increases to infinity within a proper distance of order $\sqrt{\alpha}$.
It is impossible to faithfully present the geometry through an embedding in flat 3D space.
The graph presents the change in the radius $r$, 
while the proper distance is represented by the projection on the vertical axis.
}
\label{fig:near-neck-vacuum}
\end{center}
\end{figure}

The proper distance from a point deeper inside the neck
to the starting point $r_1$ where the approximation \eqref{eq-delta-rhop} is valid
is estimated as
\begin{equation}
\Delta s \simeq \int_{r_1}^r dr' \, f_0^{1/2} r^{\prime 3/2} e^{- r^{\prime 2}/\alpha} 
\simeq \frac{\alpha}{2} \sqrt{f_0} \left[ \sqrt{a}\, e^{-a^2/\alpha} - \sqrt{r}\, e^{-r^2/\alpha} \right]
\leq \frac{\alpha}{2} \sqrt{f_0 a}\, e^{-a^2/\alpha} < \frac{\sqrt{\alpha}}{2},
\end{equation}
where we have used the condition \eqref{cond-f0}.
In other words,
the radius $r$ increases from $a$ to $\infty$ 
within the proper distance of order $\sqrt{\alpha}$.
Schematically,
the near-neck geometry looks like Fig.\ref{fig:near-neck-vacuum}.
Since the near-neck region approximated by the metric \eqref{metric-wormhole}
also has a proper length of order ${\cal O}(\sqrt{\alpha})$ \eqref{near-neck-length},
the maximal length between the neck and the surface of the star
is only of the order of $\sqrt{\alpha}$.

The energy density $-\langle T^0{}_0 \rangle$ for the vacuum in this region 
\begin{equation}
 - \langle T^0{}_0 \rangle \simeq - \frac{4}{\kappa \alpha f_0 r^3} e^{2 r^2/\alpha} 
\label{T00-vac-1}
\end{equation}
grows exponentially with $r$ from 
${\cal O}(a^{-2} \ell_p^{-2})$ at $r = r_1$ 
to infinity as $r \rightarrow \infty$.
(Recall that $r_1$ is the radius of the surface below which
the approximation in this subsection is valid.)
Note that
the energy density at $r = r_1$ is much smaller than the Planck scale
due to \eqref{cond-f0}.
But in order for the density to be sub-Planckian at $r_s$ as well,
we need
the surface radius $r_s$ of the star to satisfy
\begin{equation}
r_s - r_1 < {\cal O}(\frac{\alpha}{2a}\log(a/\ell_p)). 
\end{equation}

For the sake of curiosity,
in the case when there is no star under the neck,
the contribution of the vacuum energy density \eqref{T00-vac-1}
to the total mass is negative and of the same order as
the total mass of the black hole;
\begin{equation}
 \Delta M \simeq - \int d^3 x \sqrt{-g}\, \langle T^0{}_0\rangle 
 \simeq - 4\pi \int_a^\infty dr \frac{4 r \tilde c_0^{1/2}}{\kappa \alpha f_0^{1/2}} 
 \simeq - \frac{8\pi a}{\kappa} \sim \mathcal O(a \ell_p^{-2}) \ . 
\end{equation}
Therefore,
the vacuum contribution to mass is always smaller than that of matter.

Let us summarize the geometrical features of the space outside a static star
with spherical symmetry when the back reaction of the vacuum energy is taken into account.
While the space is foliated by 2-spheres,
the radius of the 2-sphere decreases as one moves towards the star from distance,
until one reaches the neck of the wormhole-like geometry at $r=a$,
which is a local minimum of $r$.
The radius starts to increase behind the point $r=a$ as we move further towards the origin,
until we reach the surface of the star.
In the hypothetical case 
when there is no star and the vacuum energy is the only source of gravity,
there must be a singularity at the ``center'' 
(the limit $r \to \infty$ under the neck --- see eq.\eqref{divergence-R}).
The more physical situation is that
there is a star with positive mass and an outer surface of finite radius $r_s$.
The geometry discussed above should be limited to the region outside the outer surface of the star.
The proper distance between the neck and the surface of the star is of order $\sqrt{\alpha}$ or less.

\begin{figure}
\vskip1cm
\begin{center}
\includegraphics[scale=0.5,bb=0 0 375 280]{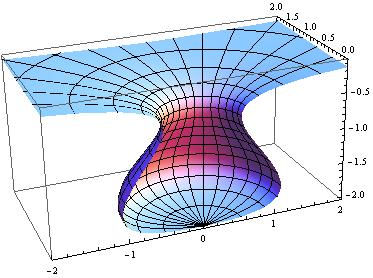}
\caption{\small 
The schematic cross-section of a black hole,
including a star under the neck.
The space outside the ``bag'' is asymptotically Schwarzschild.
While $r$ increases as we go deeper down in the vacuum,
it decreases inside the star to zero at the origin.
The notion of proper distance is not properly represented in the graph.
}
\label{fig:bag}
\end{center}
\end{figure}

In the following,
we will study
the geometry inside the star.
The radius of the 2-sphere is expected to decrease again 
after passing through the surface of the star,
until the radius goes to zero at the origin. 
Schematically,
the geometry is depicted in Fig.~\ref{fig:bag},
which is reminiscent of Wheeler's ``bag of gold'' \cite{Wheeler-Bag-Of-Gold}.
Notice that this geometry is consistent with the numerical simulation of Ref.\cite{Parentani:1994ij}
for a dynamical black hole for the same model of vacuum energy,
including Hawking radiation.
We will discuss this interior geometry in the next section.


\section{Thin shell}\label{sec:shell}

In this section, we consider the toy model of a star 
consisting of a static spherical thin shell. 
The space inside the thin shell is assumed to be Minkowski space,
which has zero vacuum energy.
The matter distribution of the shell is given by a delta function, 
and the geometry is obtained by matching across the thin shell
the external geometry of vacuum discussed in the previous section 
with the flat spacetime inside the shell.

Patching two geometries on the two sides of a thin shell imposes junction conditions.
The first junction condition is 
that the induced metric on the shell must be identical
for the bulk metrics on both sides of the shell.
The metric is expressed in the general form
\begin{equation}
ds^2 = - C(r_*) [ dt^2 - dr_*^2 ] + r^2(r_*) d\Omega^2 \ , 
\end{equation}
where $C$ and $r$ outside the shell are given by the solution 
we have discussed in the previous section. 
For flat spacetime inside the shell, 
$C$ and $r$ are given by a constant and linear function of $r_*$, respectively. 
The continuity condition of the metric
for a shell at $r_* = \rshell_*$ 
and the regularity condition at $r_0$ then determine 
$C$ and $r$ in flat spacetime ($r_* \leq \rshell_*$) to be
\begin{align}
 C(r_*) &= C(\rshell_*) \equiv \Cshell \ ,
\\
 r(r_*) &= \sqrt{\Cshell} \, (r_* - \rshell_*) + \rshell.
\end{align}
Here, the radius of the shell is denoted as $\rshell$. 
The function $C$ has discontinuity in its first derivative,
and the second derivative of $C$ involves a delta function.

The other junction condition is that 
the energy-momentum tensor on the shell 
must match with the discontinuity in the Einstein tensor 
to satisfy the Einstein equation. 
The energy-momentum tensor on the shell contributes 
to delta-function terms in the full energy-momentum tensor $T_{\mu\nu}$:
\begin{equation}
 T_{\mu\nu} = T^{\text{delta}}_{\mu\nu} + \mbox{regular terms} \ .
\end{equation}
The first term
\begin{equation}
 T^\text{delta}_{\mu\nu} 
 = \delta(\ell) S_{\mu\nu}
 \label{T=deltaS}
\end{equation}
involves a delta function of $\ell$,
which is a function of $r_*$ 
defined such that the position of the shell is at $\ell = 0$ 
and that
it gives the normal vector of the shell by 
\begin{equation}
 n_\mu = \partial_\mu \ell \ , 
\end{equation}
with $n_\mu$ normalized as $n^2 = 1$. 
(Basically, $\ell$ is the distance function from the shell.)
According to Einstein's equations,
the tensor $S_{\mu\nu}$ in eq.\eqref{T=deltaS}
can be expressed in terms of 
the discontinuity of the extrinsic curvature as
\begin{equation}
 S_{\mu\nu} = 
 \frac{1}{\kappa} \left[
 \lim_{\ell\to +0}\left(\gamma_{\mu\nu} K - K_{\mu\nu}\right)
 - \lim_{\ell\to -0}\left(\gamma_{\mu\nu} K - K_{\mu\nu}\right)
 \right] \ ,
\end{equation}
where $\gamma_{\mu\nu}$ is the induced metric on the shell, 
and $K_{\mu\nu}$, $K$ are the extrinsic curvature and its trace, respectively. 
Not only the thin shell but also the vacuum energy
contribute to the tensor $S_{\mu\nu}$.
Since there is a discontinuity in the curvature, 
the vacuum energy-momentum tensor \eqref{VEMuu}-\eqref{VEMuv} 
also contains delta-function terms. 

We identify the delta-function terms in the energy-momentum tensor as
\begin{align}
T^\text{delta}_{tt} &= \delta(\ell) \frac{\Cshell^{1/2}}{\kappa\rshell} 
 \left[2 \Cshell^{1/2} - 2\frac{dr}{dr_*}(\rshell_*)\right] \ ,
\label{Ttotaltt}
\\
T^\text{delta}_{r_* r_*} &= 0 \ ,
\label{Ttotalrr}
\\
T^\text{delta}_{\theta\theta} &= \delta(\ell) \frac{\rshell}{\kappa \Cshell^{1/2}} 
\left[ 2 \frac{dr}{dr_*}(\rshell_*) - 2 \Cshell^{1/2} 
 + \frac{\rshell}{\Cshell}\frac{dC}{dr_*}(\rshell_*) \right] \ ,
\label{Ttotalthth}
\\
T^\text{delta}_{\phi\phi} &= T^\text{delta}_{\theta\theta} \sin^{2}(\theta) \ . 
\label{Ttotalphph}
\end{align}
There is necessarily pressure in the tangential directions to support
the thin shell from collapsing. 

In order to obtain the energy-momentum tensor of the shell itself, 
the energy-momentum tensor of the vacuum should be subtracted from 
\eqref{Ttotaltt}-\eqref{Ttotalphph}.
By substituting the discontinuity of the second derivative of $C$ 
into \eqref{VEM}-\eqref{VEMuv}, 
we find that 
the delta-function term in the vacuum energy-momentum tensor is
\begin{equation}
 T^\text{vac}_{tt} 
 = 
 \frac{\alpha}{\kappa{\rshell}^2 \Cshell^{1/2}} (\partial_{r_*}\! C(\rshell_*)) \delta(\ell) \ , 
\end{equation}
while
$ T^\text{vac}_{rr} 
 = 
 T^\text{vac}_{\theta \theta} 
 = 
 T^\text{vac}_{\phi\phi} 
 = 0$.
By subtracting this vacuum energy-momentum tensor from the total energy-momentum tensor,
we find the energy-momentum tensor of the matter shell:
\begin{align}
T^\text{shell}_{tt} &= \delta(\ell) \frac{\Cshell^{1/2}}{\kappa\rshell} 
 \left[2 \Cshell^{1/2} - 2\frac{dr}{dr_*}(\rshell_*) - \frac{\alpha}{\rshell \Cshell} 
 (\partial_{r_*}\! C(\rshell_*))\right] \ ,
\label{Tshelltt}
\\
T^\text{shell}_{r_* r_*} &= 0 \ ,
\label{Tshellrr}
\\
T^\text{shell}_{\theta\theta} &= \delta(\ell) \frac{\rshell}{\kappa \Cshell^{1/2}} 
\left[ 2 \frac{dr}{dr_*}(\rshell_*) - 2 \Cshell^{1/2} 
 + \frac{\rshell}{\Cshell}\frac{dC}{dr_*}(\rshell_*) \right] \ ,
\label{Tshellthth}
\\
T^\text{shell}_{\phi\phi} &= T^\text{shell}_{\theta\theta} \sin^{2}(\theta) \ . 
\label{Tshellphph}
\end{align}

Depending on the location of the shell,
the metric functions $C$ and $r$ have different analytic approximations,
so we discuss different situations separately in the following.
First, if the shell is located well outside the Schwarzschild radius,
the geometry is well approximated by the Schwarzschild solution,
and this situation is already well understood.
In the following,
we consider two cases:
(1) the case with the shell very close to the neck
so that \eqref{cond-r1} is satisfied
and the metric is approximately given by \eqref{metric-near},
and
(2) the case when the shell is deeper down the neck
so that \eqref{cond-r1} is violated
and the metric is approximately described by \eqref{DeepInside}.

\subsection{Shell close to the neck}

Consider a shell close to the neck of the wormhole-like geometry,
i.e., $\rshell\sim a$
and the condition \eqref{cond-r1} is satisfied.
In this case,
the metric is approximately given by eq.\eqref{metric-near},
from which one can read off the functions $C(r_*)$ and $r(r_*)$.
Substituting this solution into eqs.\eqref{Tshelltt}-\eqref{Tshellphph},
we find the energy-momentum tensor of the shell:
\begin{align}
T^\text{shell}_{tt} &= \delta(\ell) 
 \frac{2 \bar c_0 \alpha}{\kappa a^4} 
 \left(a-\sqrt{\alpha}\right) 
 \left[ 1 + \frac{2 \bar c_0^{1/2}}{a} \left(r_* - a_*\right) + \mathcal O((r_* - a_*)^2)\right] 
 \ ,
\\
T^\text{shell}_{r_* r_*} &= 0 \ ,
\\
T^\text{shell}_{\theta\theta} &= \delta(\ell) 
 \frac{2a}{\kappa\sqrt{\alpha}} \left[ \left(a-\sqrt{\alpha}\right) 
 - \frac{\bar c_0^{1/2} (a^2 - \alpha)}{a^2} \left(r_* - a_*\right) 
 + \mathcal O((r_* - a_*)^2)\right] \ ,
\\
T^\text{shell}_{\phi\phi} &= T^\text{shell}_{\theta\theta} \sin^{2}(\theta) \ . 
\end{align}
Here, we have introduced a constant $\bar c_0$ by 
\begin{equation}
 c_0 = \bar c_0 \frac{\alpha}{a^2} \ ,
\end{equation}
which is of order 1 since $c_0 = \mathcal O (\alpha a^{-2})$.

The energy density $m_0$ and the (angular) pressure $P$ on the shell are found to be
\begin{align}
 m_0 &= C^{-1} T^\text{shell}_{tt} = \delta(\ell) \frac{2}{\kappa a^2} 
 \left(a-\sqrt{\alpha}\right) + \mathcal O((r_* - a_*)^2) \ ,
\label{m0near}
\\
 P &= r^{-2} T^\text{shell}_{\theta \theta} 
\notag\\
 &= 
 \delta(\ell) 
 \frac{2}{\kappa a \sqrt{\alpha}} 
 \left[\left(a-\sqrt{\alpha}\right) 
 - \frac{\bar c_0^{1/2} (a^2 - \alpha)}{a^2} \left(r_* - a_*\right) 
 + \mathcal O((r_* - a_*)^2) \right] \ . 
\label{Pnear}
\end{align}
The density $m_0$ is almost the same as the classical case
without vacuum energy for a thin shell at the Schwarzschild radius.
The quantum Schwarzschild radius $a$,
which is defined as the radius of the neck of the wormhole, 
is only slightly larger than the classical Schwarzschild radius $a_0$.

In terms of the Planck length $\ell_p$,
the mass density $m_0$ and the pressure $P$ behave as 
\begin{align}
 m_0 &\sim \mathcal O(a^{-1}\ell_p^{-2}) \ , 
\\
 P &\sim \mathcal O(\ell_p^{-2}\alpha^{-1/2}) \ . 
\end{align}
The surface energy density is of the order of $a^{-1} \ell_p^{-2}$ as expected,
as the total mass is of order $a/\kappa$ and the area of order $a^{2}$.
On the other hand, the pressure is very large
in order to support the shell against gravity from collapsing.
In fact,
it diverges in the classical case
as the shell gets too close to the Schwarzschild radius.
By taking into account the quantum effects,
the pressure is regularized, 
but is close to the Planck scale unless $N \gg 1$.

\subsection{Shell deep under the neck}

Next, we consider the scenario when the shell is deeper inside the wormhole
(that is, the condition \eqref{cond-r1}) is violated),
so that the metric is approximated by eq.\eqref{DeepInside}. 
Then the energy-momentum tensor of the shell can be calculated as 
\begin{align}
T^\text{shell}_{tt} &= \delta(\ell) 
 \frac{2}{\kappa r} \Cshell \left[1 
 - \frac{1}{f_0^{1/2}\,r^{3/2}} e^{r^2/\alpha} \left(1 + \frac{\alpha}{16 r^2} 
 + \mathcal O(\alpha^2)\right)\right] \ ,
\\
T^\text{shell}_{r_* r_*} &= 0 \ ,
\\
T^\text{shell}_{\theta\theta} &= \delta(\ell) 
 \frac{r^2}{\kappa}\left[- 2 r 
 + \frac{4 r^{3/2}}{\alpha f_0^{1/2}} e^{r^2/\alpha} 
 \left(1 - \frac{3 \alpha}{16 r^2} + \mathcal O(\alpha^2)\right)\right] \ ,
\\
T^\text{shell}_{\phi\phi} &= T^\text{shell}_{\theta\theta} \sin^{2}(\theta) \ . 
\end{align}

Note that the second term in $T^{\text{shell}}_{tt}$ has an exponential factor
that is sensitive to $r$ and blows up quickly for a tiny increase in $r$.
Starting with a positive energy density $m_0$ of the shell at some radius $r$
where \eqref{metric-near} is a good approximation,
we find that the energy density of the shell decreases very quickly 
if we move the shell further down the throat.
The density $m_0$ becomes negative beyond a certain point.
For even larger radius of the shell
(further deep down the throat),
the density $m_0$ is negative with a larger magnitude.
This means that, 
for a physical shell with positive energy density $m_0 > 0$,
the radius $r$ of the shell ($\rshell$) is bounded from above,
and the location of the shell from the neck is bounded from below,
that is,
it cannot be located too deep under the neck.
Due to the exponential factor $e^{-2r^2/\alpha}$,
the deviation of the shell from the neck of the wormhole-like structure
cannot be much larger than the order of $r - a \sim {\cal O}((\alpha/a)\log(a^2/\alpha))$.
That is,
a shell under the neck is always very close to the neck.


\section{Incompressible fluid}\label{sec:fluid}

As another model for a static star,
we consider incompressible fluid. 
The energy-momentum tensor for the perfect fluid is given by 
\begin{equation}
 T^m_{\mu\nu} = (m+P) u_\mu u_\nu + P g_{\mu\nu} \ , 
\end{equation}
where $m$ is the mass density, $P$ is the pressure 
and $u^\mu$ is the velocity 4-vector which is normalized as $u^2 = -1$. 
For static configurations, 
the spatial components of the velocity vector vanishes:
$u_i=0$. 
The energy-momentum tensor expressed in the $(u,v)$-coordinates is then
\begin{align}
 T^m_{uu} 
 &= 
 T^m_{vv}
 = \frac{1}{4} C (m+P) \ , 
\\
 T^m_{uv}
 &= 
 \frac{1}{4} C (m-P) \ , 
\\
 T^m_{\theta\theta} 
 &= 
 T^m_{\phi\phi} /\sin^2 \theta
 = r^2 P \ , 
\end{align}
and other components of $T^m_{\mu\nu}$ vanish.
We shall assume that 
the matter and the vacuum satisfy 
the law of energy-momentum conservation separately. 
The conservation law for the matter
is written as 
\begin{equation}
 \left[ m(r)+P(r) \right] \frac{C'(r)}{2C(r)} + P'(r) = 0 \ . 
\label{ConsFluid}
\end{equation}
We approximate the vacuum energy-momentum tensor by
that of the 2-dimensional scalar fields \eqref{VEM}-\eqref{VEMuv} with \eqref{ZeroT}. 
Since 2-dimensional energy-momentum tensor $T_{\mu\nu}^{(2D)}$ satisfies 
the 2-dimensional conservation law while the 4-dimensional energy-momentum tensor 
$T_{\mu\nu}^\Omega$ satisfies the 4-dimensional conservation law, 
it is implicitly assumed that
\begin{equation}
 T_{\theta \theta}^\Omega 
 = 
 T_{\phi\phi}^\Omega
 = 
 0 \ .
\end{equation} 

By substituting the energy-momentum tensors into 
the semi-classical Einstein equations \eqref{Einstein}, 
we obtain the following differential equations:
\begin{align}
 0 &= 
 -\frac{1}{8 r^2 C(r)^2}
 \Bigl\{2 C(r) F(r) 
 \left[
  \alpha C'(r) F'(r)+F(r) \left(\alpha  C''(r)- 2 r C'(r)\right)
 \right] 
\notag\\&\qquad\qquad\qquad
 +4 r C(r)^2 F(r) F'(r) + 2 \kappa r^2 C(r)^3 (m(r)+P(r)) 
 -3 \alpha  F(r)^2 C'(r)^2 \Bigr\}
 \ , 
\label{eq1}
\\
 0 &= 
 \frac{1}{4 r^2 C(r)^2}
  \Bigl[
  - \alpha  C(r) F(r) \left(F(r) C''(r)+C'(r) F'(r)\right) 
   -2 C(r)^2 F(r) \left(r F'(r)+F(r)\right) 
\notag\\&\qquad\qquad\qquad
   +C(r)^3 \left(2 - \kappa r^2 (m(r) - P(r))\right) 
   + \alpha  F(r)^2 C'(r)^2
  \Bigr] 
 \ , 
\label{eq2}
\end{align}
where eq.\eqref{eq1} is the $(u,u)$- and $(v,v)$-components of the Einstein equation,
and eq.\eqref{eq2} the $(u,v)$-component of the Einstein equation. 
The diagonal parts of the angular components are equivalent to
the these equations up to the conservation law of the energy-momentum tensor. 
By taking the difference of \eqref{eq1} and \eqref{eq2}, 
we obtain 
\begin{equation}
 F(r) = \sqrt{ \frac{4 C^3(r)(1 + \kappa r^2 P(r))}
 {4 C^2(r) - 4 r C(r) C'(r) + \alpha C^{\prime\,2}(r)}} \ . 
\end{equation}
Substituting this equation back to one of the two differential equations above, 
we find
\begin{align}
 0 
 &= 
 r^2 \kappa \left(m(r)+r P'(r) + 3 P(r)\right) 
\notag\\&\quad
 + r \left[-4 + 4 r^2 \kappa m(r) + 2 r^3 \kappa P'(r)+ \alpha r \kappa P'(r) 
 + 2 \kappa P(r) \left(\alpha +2 r^2\right)\right] 
 \rho '(r) 
\notag\\&\quad
 + \left[-4 r^2 - 2 \alpha + 2 r^2 \kappa m(r) \left(\alpha +2 r^2\right) 
 +3 \alpha r^3 \kappa P'(r) + 6 \alpha r^2 \kappa P(r)\right] 
 \rho^{\prime\,2}(r) 
\notag\\&\quad
 + \alpha  r \left[-2 + 4 r^2 \kappa m(r)+\alpha  r \kappa P'(r) 
 + 2 \kappa P(r) \left(\alpha +r^2\right)\right] 
 \rho^{\prime\,3}(r) 
\notag\\&\quad
 +\alpha^2 r^2 \kappa (m(r)+P(r)) 
 \rho^{\prime\,4}(r) 
\notag\\&\quad
 -2 \left(r^2 \kappa P(r)+1\right) \left(r^2-\alpha \right) \rho ''(r)
\ . \label{eq3}
\end{align}

In order to solve this equation uniquely, 
we have to impose additional conditions on the matter. 
At this point, mass density $m(r)$ and pressure $P(r)$ are independent. 
Here are infinitely many solutions to the equation above 
corresponding to perfect fluids of different kinds.%
\footnote{%
An exact solution was also found in \cite{Carballo-Rubio:2017tlh} 
though it has negative pressure (and negative mass singularity). 
Notice that solutions with negative pressure 
will be excluded in this paper (See also Sec.~\ref{sssec:num-in-c}). 
}
Here,
we consider the incompressible fluid, 
which has a constant mass density $m(r) = m_0$. 
In this case, the conservation law for the fluid \eqref{ConsFluid} is solved as 
\begin{equation}
 P(r) = - m_0 + P_0 e^{-\rho(r)} \ , 
\label{Press}
\end{equation}
where $P_0$ is an integration constant. 
Then, the differential equation \eqref{eq3} becomes 
\begin{align}
 0 &= 
 -\kappa  r^2 \left(2 m_0 e^{\rho
   (r)}-3 P_0\right) 
\notag\\&\quad
 +r \left[-2 \alpha  \kappa  m_0 e^{\rho (r)}+\kappa  P_0
   \left(2 \alpha +3 r^2\right)-4 e^{\rho (r)}\right] \rho'(r) 
\notag\\&\quad
 +\left[4 \kappa  m_0 r^2 \left(r^2-\alpha \right)
   e^{\rho (r)}+P_0 \left(5 \alpha  \kappa  r^2-2 \kappa  r^4\right)-2 \left(\alpha
   +2 r^2\right) e^{\rho (r)}\right] 
 \rho^{\prime\,2}(r) 
\notag\\&\quad
 +\alpha  r \left[2 \kappa  m_0 \left(r^2-\alpha
   \right) e^{\rho (r)}-\kappa  P_0 \left(r^2-2 \alpha \right)-2 e^{\rho
   (r)}\right] 
 \rho^{\prime\,3}(r) 
\notag\\&\quad
 + 2 \left(r^2-\alpha \right) 
 \left(\kappa  m_0 r^2 e^{\rho (r)}-\kappa  P_0 r^2-e^{\rho (r)}\right) 
 \rho ''(r) 
\ . \label{eq4}
\end{align}

The results we obtained for incompressible fluid are solutions without 
horizon or singularity at finite $r \gg \ell_p$.

\subsection{Classical divergence}\label{sec:Classical}

In order to appreciate the difference due to the vacuum energy,
we first review several facts of the classical case.
For the classical case
(i.e. when the vacuum energy $T_{\mu\nu}^\Omega$ is absent), 
the pressure is solved as
\begin{equation}
 P(r) = m_0 \frac{\sqrt{3- \kappa m_0 r^2} - \sqrt{3 - \kappa m_0 r_s^2}}
{3\sqrt{3-\kappa m_0 r_s^2} - \sqrt{3-\kappa m_0 r^2}} \ , 
\label{ClassPress}
\end{equation}
where $r_s$ is the radius at the surface of the matters. 
In order for this pressure to be positive at $r=0$, 
it must satisfy the condition
\begin{equation}
 \kappa m_0 <  \frac{8}{3 r_s^2} \ . 
\label{ClassCond}
\end{equation}
Otherwise, the denominator of \eqref{ClassPress} becomes zero at  some $r\geq 0$.
That is, 
the pressure diverges if the density of the fluid is too large 
for a given radius of the surface,
or equivalently,
if the radius of the star is too small.

From the junction condition at the surface $r_s$, 
we see that the radius of the surface $r_s$ should be not only larger than the Schwarzschild radius $a_0$,
but Buchdahl's theorem says that it should satisfy the inequality \cite{Buchdahl:1959zz}
\begin{equation}
 r_s > \frac{9}{8} a_0 \ .
 \label{CondRadius}
\end{equation}
Therefore, 
classically,
a star must collapse to a singularity
under gravitational force when the condition
\eqref{ClassCond} or \eqref{CondRadius} is violated.

\subsection{Geometry below the surface of the star}

Contrary to the classical case, 
the pressure no longer diverges if the vacuum energy is taken into account. 
Eq.~\eqref{Press} implies that 
the pressure diverges only when $C$ goes to zero. 
But one can prove that $C$ can never go to zero for $r\gg \alpha^{1/2}$, 
in a way similar to how
we have proved the absence of horizon for the vacuum solution \cite{Ho:2017joh}.

Here we sketch the proof that $C$ can never go to zero
in the presence of the incompressible fluid.
We first assume that $C=0$ at some point $r=r_d$.
This implies that $\rho\to-\infty$ as $r \to r_d$, 
and thus $\rho'$ also diverges there. 
Therefore,
in a sufficiently small neighborhood of $r = r_d$,
terms proportional to $\rho^{\prime\,3}$ in the 3rd line
dominate over the first 2 lines in eq.\eqref{eq4}. 
Furthermore,
in the coefficients of $\rho^{\prime\,3}$ and $\rho''$ in eq.\eqref{eq4},
the terms proportional to higher powers of $e^\rho$ are negligible 
in comparison with those proportional to lower powers of $e^\rho$. 
As a result, the differential equation can be approximated by 
\begin{equation}
 \alpha (r^2-2\alpha) \rho^{\prime\,3} + 2 r(r^2-\alpha) \rho'' \simeq 0 \ . 
\end{equation}
This implies that the divergence of $\rho'$ must have the form 
\begin{equation}
 \rho' \simeq \pm 
 \left(\frac{r_d(r_d^2-2\alpha)}{\alpha(r_d^2 - 2\alpha)}\right)^{1/2} (r-r_d)^{-1/2} \ . 
\label{wormhole2nd}
\end{equation}
Up to a constant factor,
this is the same solution as the vacuum solution \eqref{rho-sol}, 
for which $C$ does not go to zero.
Hence we conclude that the vacuum energy
regularizes the pressure $P$ such that
it never diverges at finite $r$.

Let us now take a step further and study the geometry below the surface of the star 
but for $r$ much larger than the Planck length, $r\gg \alpha^{1/2}$. 
While we have shown in the above that $C$ cannot vanish 
and $P$ cannot diverge but the structure of \eqref{wormhole2nd} appears instead, 
notice that it is not necessary that 
the behavior of \eqref{wormhole2nd} appears in the geometry below the surface of the star. 
It is possible that neither $C=0$, divergence of $P$ nor \eqref{wormhole2nd} is realized. 
The radius $r$ is expected to go to zero at the center of the star, 
and in fact, $r$ has one local maximum of $r$ under the neck of wormhole at $r=a$, 
as we will see below. 
It is possible that there is an additional local minimum of $r$ 
where \eqref{wormhole2nd} holds, in addition to the neck at the Schwarzschild radius. 
As we will see in the next section on numerical analyses,
this additional local minimum in $r$ appears
when the mass density $m_0$ is too small for a star with Schwarzschild radius at $r=a$. 
There would be no additional local maximum other than 
that between 2 local minima in this case, 
as long as the density $m_0$ is constant. 
Hence, we shall try to 
describe the behavior of $C(r)$ inside the fluid
in the following.


Let us now show that there is a 
local maximum of $r$ under the neck.
Since the pressure must vanish at the surface of the star, 
the pressure is negligible just below the surface. 
Assuming that the surface is not far away from the neck of the wormhole, 
$\rho'(r)$ is still very large near the surface and 
\eqref{eq4} is approximately the same as
\begin{equation}
 0 =  \alpha  r \left(\kappa m_0 r^2 -2 \right) 
 \rho^{\prime\,3}(r) 
 - 2 \left(r^2-\alpha \right) \rho ''(r) \ . 
\end{equation}
Then, $\rho'(r)$ is solved as 
\begin{equation}
 \rho'(r) 
 = 
 \pm \frac{2^{1/2}}
 {\sqrt{c - \alpha \kappa m_0 (r^2 - \alpha) + \alpha (2- \alpha\kappa m_0)\log (r^2 - \alpha)}} \ , 
\end{equation}
with an integration constant $c$. 
The argument in the square root in the denominator should be positive just below the surface, 
but approaches to zero as one moves inside, with $r$ increasing. 
When it goes to zero at $r = b$, 
$C(r)$ and $F(r)$ behave around $r=b$ as 
\begin{align}
 C(r) &\simeq c_b e^{2 \sqrt{k_b (b-r)}} \ , 
& F(r) &\simeq \pm\sqrt{ \frac{2(b-r)}{\alpha k_b}} \ , 
\end{align}
where 
\begin{equation}
 k_b = \frac{4(b^2 - \alpha)}{\alpha b (\kappa m_0 b^2 - 2)} 
 \simeq \frac{4}{\alpha \kappa m_0 b} \ ,  
\end{equation}
and the sign on $F$ is negative outside the local minimum $r=b$ and positive inside it. 
This implies that the radius $r$ has a local maximum $r=b$, 
and starts to decrease again as one goes down towards the center of the star from $r=b$. 

As one goes further inside this geometry, 
$\rho$ would continue to decrease, and hence, 
the terms with $e^{\rho}$ would be negligible, namely, 
\begin{equation}
 m_0 e^{\rho(r)} \ll P_0 \ , 
\end{equation}
for $r\ll r_s$. 
Then, the leading order terms of \eqref{eq4} give the differential equation
\begin{equation}
 0 = -3 -3 r \rho'(r) + 2 r^2 \rho^{\prime\,2}(r) - \alpha r \rho^{\prime\,3}(r) + 2 r^2 \rho''(r) \ . 
\label{eq6}
\end{equation}
If $\rho(r) > \mathcal O(\alpha^0)$, 
a solution of \eqref{eq6} is given by 
\begin{equation}
 \rho(r) = - \frac{r^2}{\alpha} + \mathcal O(\alpha^0) \ . 
\label{rho_dec}
\end{equation}
However, this is not appropriate for a solution inside the fluid 
since $\rho(r)$ should be increasing function
with respect to $r$,
at least near the surface $r\lesssim r_s$. 
Therefore, $\rho(r)$ should behave as $\rho(r) = \mathcal O(\alpha^0)$. 
Then, \eqref{eq6} is approximated as 
\begin{equation}
 0 = -3 -3 r \rho'(r) + 2 r^2 \rho^{\prime\,2}(r) + 2 r^2 \rho''(r) \ , 
\end{equation}
which is solved as 
\begin{equation}
 \rho(r) = - \frac{1}{2} \log r + \log \left(r^{7/2} + c_1^\text{in}\right) 
 + \frac{1}{2} \log c_0^\text{in} \ , 
\label{rho_inside}
\end{equation}
where $c_0^\text{in}$ and $c_1^\text{in}$ are integration constants. 

If $c_1^\text{in}$ is negative, $C(r)$ approaches 
to zero around $r \sim |c_1^\text{in}|^{2/7}$, 
and then 
the solution should be connected to the wormhole-like structure given by \eqref{wormhole2nd}, 
which corresponds to the case with too small density. 
If $c_1^\text{in}$ is positive, $\rho(r)$ becomes a decreasing function 
for $r < 6^{-2/6} (c_1^\text{in})^{2/7}$. 
In this case, if
$\rho(r)$ becomes as large as that at $r_s$, 
the approximation \eqref{eq6} is no longer good, 
but the pressure becomes zero when $\rho(r) = \rho(r_s)$. 
As we will see in the next section, 
the pressure becomes 
zero at
$r = r_s^{inner} < r_s$, 
which corresponds to the case in which the density $m_0$ is too large 
for a star with Schwarzschild radius $a$. 
If $c_1^\text{in}$ is positive and sufficiently small, 
pressure $P$ does not goes to zero for $\alpha^{1/2} \ll r < r_s$, 
which would be a physical configuration of 
a black hole with back reaction from vacuum energy. 

The constant $c_0^\text{in}$ is related to the overall normalization 
of $C(r)$ and determined by the junction condition to 
a solution for $\rho(r) \sim \rho(r_s)$. 
The differential equation \eqref{eq6} is good 
if $e^{\rho(r)-\rho(r_s)}$ is sufficiently small 
even if it is not so small as $\alpha$, namely,
eq.\eqref{eq6} is good
for $e^{\rho(r)-\rho(r_s)} = \mathcal O(\alpha^0)$ but with $e^{\rho(r)-\rho(r_s)} \ll 1$. 
In this case, \eqref{rho_inside} implies that the pressure is 
of the same order to the density, at least 
if $r$ is sufficiently larger than the Planck length. 
Therefore, the gravitational collapse due to strong pressure 
would not happen in this model. 



Finally, we consider expansion around $r=0$. 
Assuming that $\rho$ is expanded as 
\begin{equation}
 \rho(r) = \rho_0 + \rho_1 r + \rho_2 r^2 + \rho_3 r^3 + \rho_4 r^4 + \cdots \ , 
\end{equation}
\eqref{eq4} is solved as 
\begin{align}
 \rho_3 
 &= 
 \frac{\rho_1}{6 \alpha} 
 \left[
  1 + 3 \alpha \rho_1^2 + 
  \alpha \kappa \left(m_0 - e^{-\rho_0} \right)P_0 \left(1 + 2 \alpha \rho_1^2 \right)
 \right] \ , 
\\
 \rho_4 
 &= 
 \frac{1}{48 \alpha} 
 \left(1 + 2 \alpha \rho_1^2 \right) 
 \left[
  2 \left(m_0 + 7 \rho_1^2 + 8 \alpha \kappa m_0 \rho_1^2\right) 
  - 3 \kappa P_0 \left(1 + 4 \alpha \rho_1^2\right) 
 \right] \ .  
\end{align}
In order not to have the conical singularity at $r=0$, 
the metric should satisfy $g_{rr}=1$ at $r=0$. 
Since we have 
\begin{equation}
 g_{rr}(r=0) = \frac{C(r=0)}{F^2(r=0)} = 1 + 2 \alpha \rho_1^2 \ , 
\end{equation}
the absence of the conical singularity at $r=0$ implies 
\begin{equation}
 \rho_1 = 0 \ . 
\end{equation}
In this case, $\rho(r)$ is expanded around $r=0$ as 
\begin{align}
 \rho(r) &= \rho_0 + \frac{1}{48 \alpha} \left(2 m_0 - 3 e^{-\rho_0} P_0\right) r^4 + \cdots 
\notag\\
 &= \rho_0 - \frac{1}{48 \alpha} \left(m_0 + 3 P(r=0)\right) r^4 + \cdots \ . 
\end{align}
Since the density $m_0$ and the pressure $P(r)$ should be positive at $r=0$, 
$\rho$ is a decreasing function
(with respect to $r$)
around $r=0$. 
This implies that the negative vacuum energy becomes larger than 
the density of the incompressible fluid. 
This behavior is consistent with eq.\eqref{rho_inside}, 
which becomes a decreasing function for small $r$. 
It should be noted, however, that the behavior of $\rho$
in $r \lesssim \alpha^{1/2}$ is determined by physics at the Planck scale 
and the semi-classical Einstein equation cannot give a good prediction in this scale. 
The argument here, about the expansion around $r=0$, 
for the sake of completeness,
is just to show that the behavior of \eqref{rho_inside}, 
as a decreasing function for small $r$, 
is consistent with the semi-classical Einstein equation around $r=0$. 


\section{Numerical results}\label{sec:numerical}

In order to see more detailed structure of the geometry
for the star of incompressible fluid, 
we resort to numerical methods.
Since $C$ and $F$ are not single-valued functions of $r$, 
it is convenient to use the tortoise coordinate $r_*$ 
defined in \eqref{tortoise}
to see the whole structure of the solution. 
The semi-classical Einstein equation gives the following two differential equations 
\begin{align}
 0 &= \kappa  P_0 \, r^2(r_*) e^{\rho(r_*)} 
 + 2 r(r_*) r''(r_*) 
\notag\\&\qquad
 -4 r(r_*) r'(r_*) \rho'(r_*) + 2 \alpha \rho''(r_*) - 2 \alpha \rho^{\prime\,2}(r_*) \ ,
 \label{neq1}
\\
 0 &= 
 2 e^{2 \rho(r_*)}
 + \kappa \left(P_0 e^{\rho(r_*)} - 2 m_0 e^{2\rho(r_*)}\right) r^2(r_*)  
\notag\\&\qquad
 -2 r(r_*) r''(r_*) -2 r^{\prime\,2}(r_*) - 2 \alpha \rho''(r_*) 
 \label{neq2}
\end{align}
for the two functions $\rho(r_*)$ and $r(r_*)$.
(A prime refers to the derivative with respect to $r_*$.)
Here, we have used $m(r_*) = m_0$ and \eqref{Press}. 

Since we are considering the Boulware vacuum,
there is no incoming or outgoing energy at the spatial infinity $r\to\infty$.
The quantum effects are expected to approach to zero at large distance,
and the geometry should be an asymptotically Schwarzschild space. 
Hence the boundary conditions for $\rho(r_*)$ and $r(r_*)$ is that
they are approximated by the Schwarzschild solution
with a given classical Schwarzschild radius $r=a_0$ at large $r_*$. 

With the asymptotically Schwarzschild boundary condition,
one can solve the Einstein equation for $m=P=0$ at large $r_*$,
and then turn on the energy density $m$ and the pressure $P$
at the radius $r = r_s$
(or in terms of the tortoise coordinate, $r_* = r_{*s}$)
of the surface of the perfect fluid.
Since the pressure must vanish at the surface of the star,
the parameter $P_0$ is fixed by the condition $P(r_{*s})=0$.

Now we have 3 physical parameters $a_0$, $m_0$, $r_{*s}$ 
and 2 coupling constants $\alpha$, $\kappa$. 
The constant $\kappa$ can be absorbed by 
redefinition of $m_0$ and $P_0$,
so effectively there is only one coupling constant $\alpha$ to tune in the numerical simulation.
If there is no additional source of energy other than the fluid and the vacuum,
one of the 3 physical parameters ($a_0$, $m_0$, $r_{*s}$) should be fixed by the other 2 parameters. 
For example, 
if we fix the energy density $m_0$ and the surface radius $r_{*s}$ of the incompressible fluid,
we have then fixed the total mass of the star,
and one would expect that the classical Schwarzschild radius $a_0$, 
which is also related to the total mass of the system, should be consequently uniquely fixed. 

Mathematically, 
on the other hand,
the differential equations \eqref{neq1} and \eqref{neq2}
can be solved at least locally for arbitrary values of the 3 parameters $a_0$, $m_0$, $r_{*s}$.
As we solve the Einstein equation with the boundary condition imposed at large $r_*$,
its solution at smaller $r_*$ is determined for arbitrary choice of the three parameters.
The relation among the 3 parameters resides in the regularity condition at smaller $r_*$.
In general,
the solution involves a singularity, 
unless the 3 parameters are chosen suitably.
Physically we can interpret it as the effect of the presence of other sources of energy.
That is, an additional mass to compensate the difference between the Schwarzschild mass $a_0$
and the mass computed from $m_0$ and $r_{*s}$. 
Since the Einstein equations we use do not contain the energy of any other sources,
this additional energy source must appear as a singularity.
The geometry must have a singularity as a source of energy to cover the mismatch.
(Recall that the Schwarzschild solution for a star of mass $a_0/2$
is in fact a vacuum solution with a singularity at the origin.)

In principle, the relation between 3 parameters $m_0$, $r_{*s}$ and $a_0$ can be derived by 
comparing the classical Schwarzschild radius $a_0$ and the total mass calculated from $m_0$ and $r_s$. 
However, in order to calculate the total mass from $m_0$ and $r_{*s}$, 
we need to know details on the geometry. 
Therefore, we would not know beforehand 
the suitable value of $m_0$ for given $a_0$ and $r_{*s}$,
until we obtain the solution for the geometry. 
That is, even though conceptually
$m_0$ can be viewed as a function
$m_0 = \hat{m}_0(a_0, r_{*s})$ of $a_0$ and $r_{*s}$,
numerically,
this function can only be found by trial and error, 
namely, by solving the differential equation 
for diverse values of $m_0$ for fixed $a_0$ and $r_{*s}$.

Therefore,
in our numerical calculation,
all 3 physical parameters $a_0$, $r_{*s}$ and $m_0$
are treated as free parameters
that we can fix arbitrarily by hand.
The value of $a_0$ has to be fixed first when we impose
the asymptotically Schwarzschild boundary conditions on $\rho(r_*)$ and $r(r_*)$ at large $r_*$.
Then the value of $r_{*s}$ needs to be fixed by hand to determine
where to turn on the energy-momentum tensor of the incompressible fluid.
After that, 
a range of different values of $m_0$ are explored.
In general, 
an arbitrary choice of $m_0$ would produce a singularity at small $r_*$.
We know that we have the suitable value $m_0 = \hat{m}_0(a_0, r_{*s})$ 
only when the geometry is found to be regular everywhere.
If the density is too small, i.e. $m_0 < \hat{m}_0(a_0, r_{*s})$, 
the singularity should have positive mass. 
The vacuum solution with singularity \eqref{divergence-R} in the limit $r\to\infty$
is an example of the special case for $m_0=0$,
which is obviously too small for a non-zero $a$. 
Similarly,
if the density is too large,
i.e. $m_0 > \hat{m}_0(a_0, r_{*s})$,
the singularity should have a negative mass.

\subsection{Star inside the Schwarzschild radius}

We first consider those stars whose surfaces lie inside the neck of the wormhole-like structure. 
Since $r=a$ is a minimum of the radius, 
we have the radius of the surface of the star $r_s > a$.
We shall label the tortoise coordinate $r_*$ for the star's surface at $r = r_s$ as $r_* = r_{*s}$,
and that for $r = a$ as $r_* = a_*$.
That is, $r(r_{*s}) = r_s$ and $r(a_*) = a$. 
In our numerical simulations,
we take $\alpha=0.05$ and $a_0=10$. 
This corresponds to a relatively small star but 
it is more convenient to emphasize the quantum effects. 
Although the correction from the quantum effects would be 
quantitatively much smaller for a black-hole-like object in the real world, 
the qualitative behavior is expected to be the same. 

The neck of the wormhole is then found to be located at $r_* = a_* \simeq -56.685$,
where the radius is $r = a \simeq 10.0190$ approximately.
We put the surface of the star below the neck at $r_{*s} = - 60$,
where $r_s = r(r_{*s}) \simeq 10.0191$. 
Numerical simulations are then carried out
for different values of $m_0$ and 
we have obtained 3 types of geometries depending on the value of $m_0$.

\subsubsection{Too small density: $m_0 \ll \hat{m}_0(a_0, r_{*s})$}

\begin{figure}
\begin{center}
\includegraphics[scale=0.7,bb=0 0 259 170]{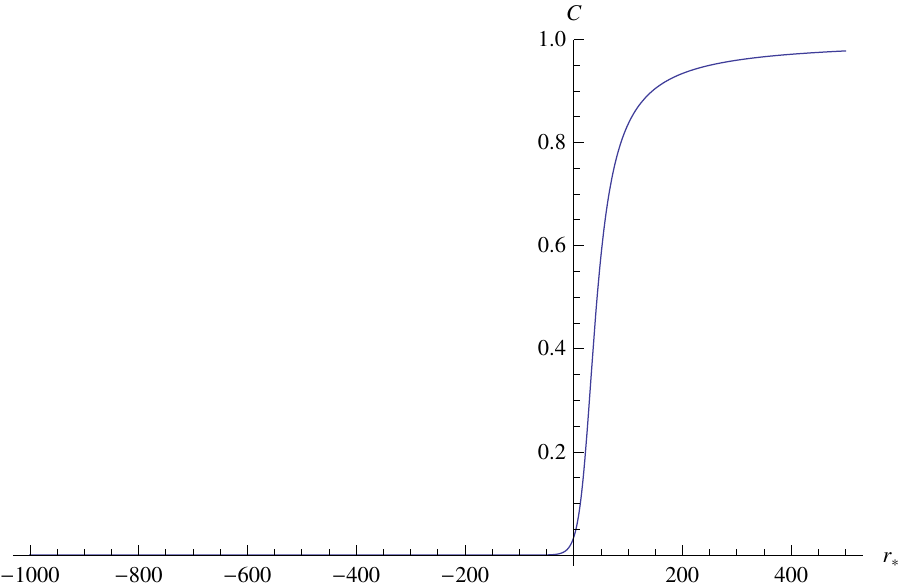}
\hspace{24pt}
\includegraphics[scale=0.7,bb=0 0 259 170]{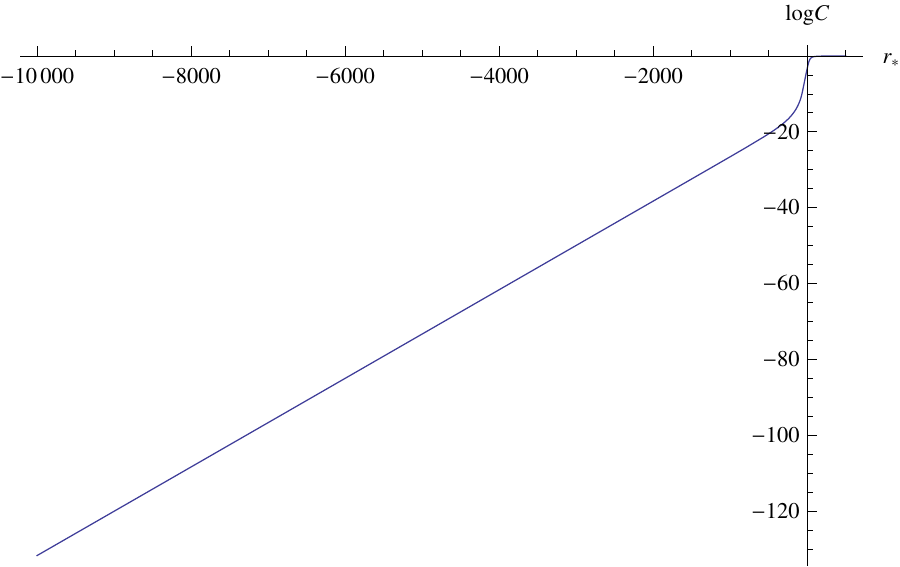}
\caption{\small 
The factor $C$ for $\kappa m_0 = 25$, $r_{*s}=-60 < a_*$, $a_0=10$ and $\alpha=0.05$. 
The plot on the left shows that
$C$ becomes very small around and inside the neck.
The plot on the right shows that the behavior of $C$ at large negative $r_*$
resembles that of the vacuum solution described in Sec.~\ref{sec:behindtheneck},
with $\log C = 2\rho$ proportional to $r_* (\propto r^2)$.
}
\label{fig:c-s}
\end{center}
\end{figure}

\begin{figure}
\begin{center}
\includegraphics[scale=0.7,bb=0 0 259 170]{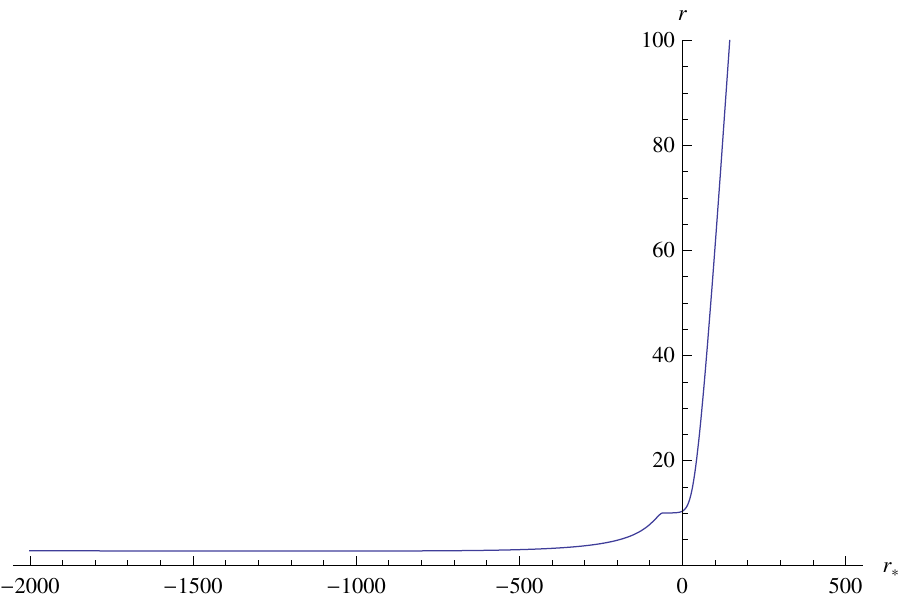}
\hspace{24pt}
\includegraphics[scale=0.7,bb=0 0 259 170]{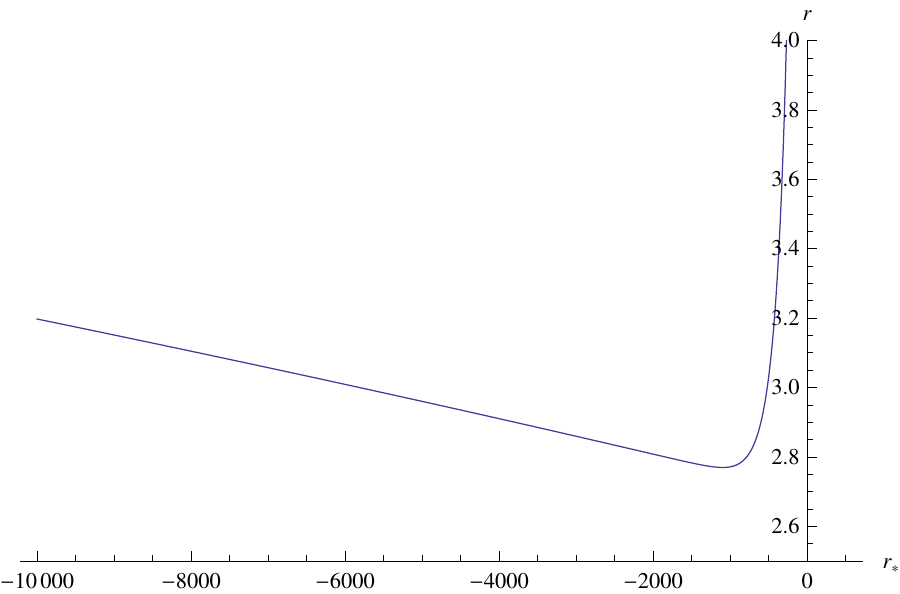}
\caption{\small 
The radius $r$ for $\kappa m_0 = 25$, $r_{*s}=-60 < a_*$, $a_0=10$ and $\alpha=0.05$. 
On the left:
The radius $r$ decreases as $r_*$ decreases for $r_*> a_*$,
but starts to increase for $r_* \lesssim a_*$. 
It starts to decrease again from $r_* \simeq - 60.003$, 
which is slightly inside the surface $r_{*s}$. 
On the right:
The radius $r$ starts to increase again with decreasing $r_*$
for $r_* < - 1097$.
}
\vskip1cm
\label{fig:r-s}
\end{center}
\end{figure}

\begin{figure}
\begin{center}
\includegraphics*[scale=0.08,bb=0 300 1430 2300]{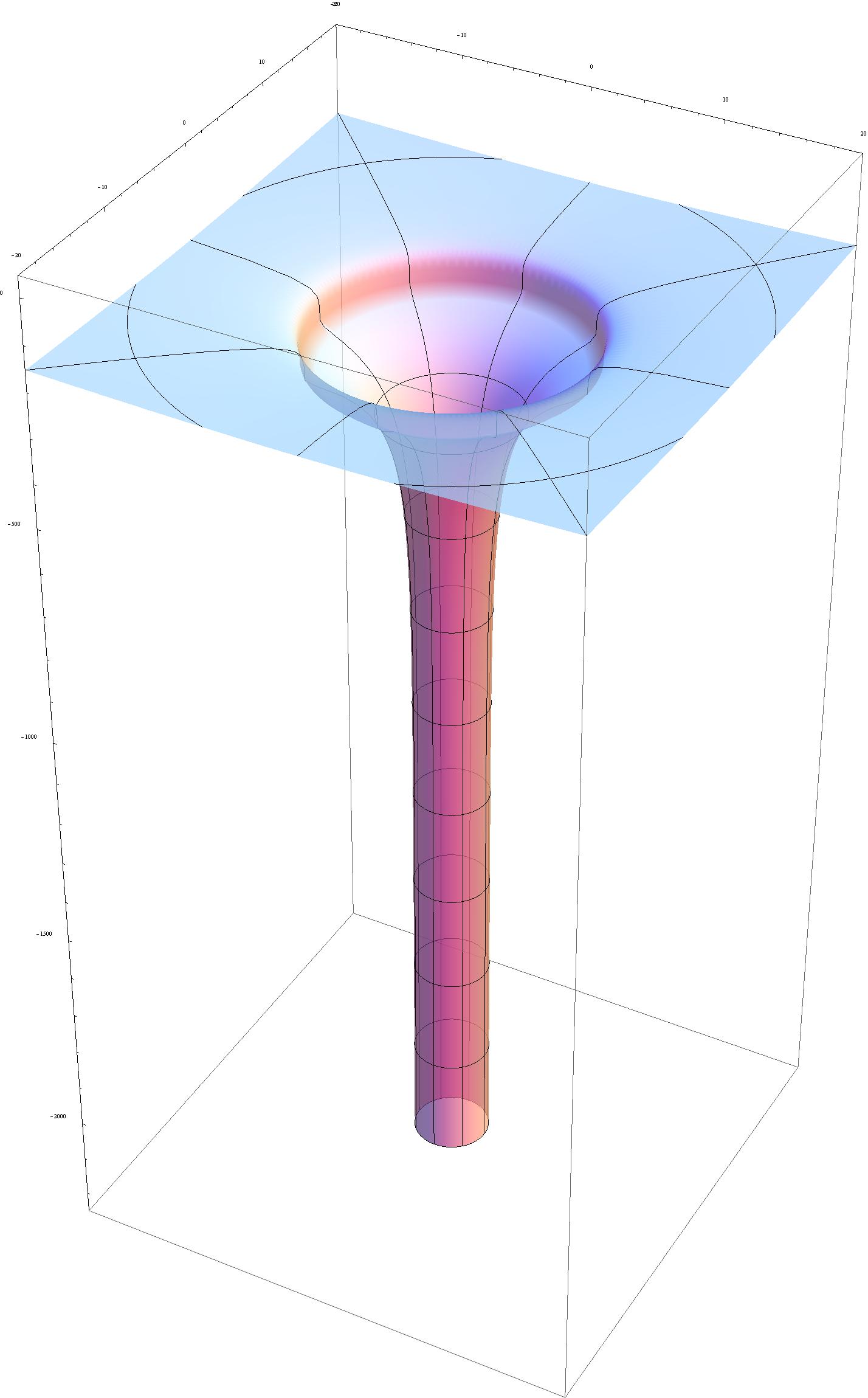}
\hspace{2cm}
\includegraphics*[scale=0.08,bb=0 300 1430 2300]{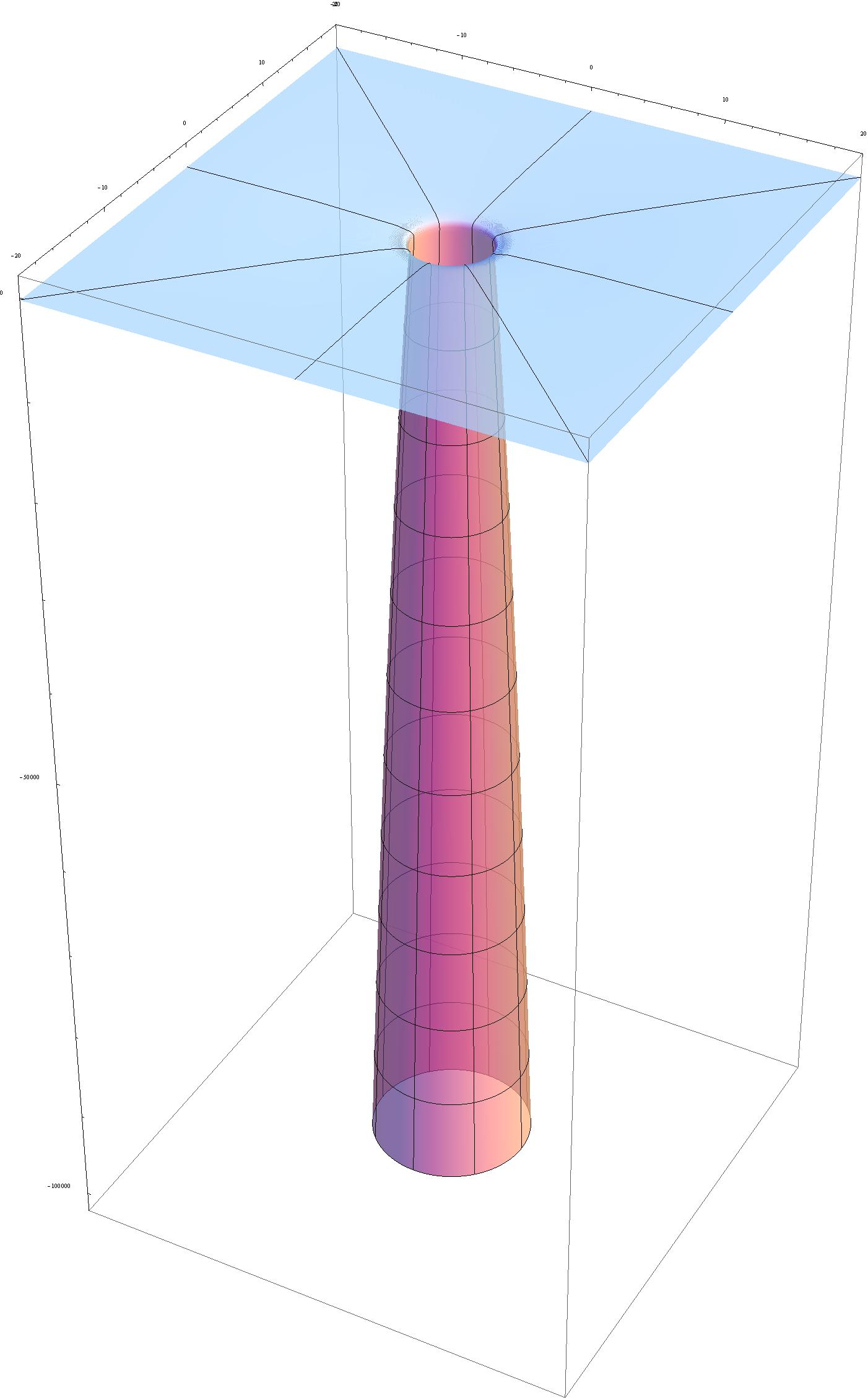}
\caption{\small 
Schematic 3D picture for the case of Fig.\ref{fig:r-s}.
The radius $r$ corresponds to that in the horizontal direction while 
the distance along the 2D surface
in the picture
is given by the change in the $r_*$ coordinate. 
Both pictures show the radius $r$ on the same scale,
but the range of $r_*$ shown in the picture on the right
is much larger than that on the left.
The radius $r \to \infty$ in both limits $r_* \to \pm \infty$.
Notice that the distance along the 2D surface is different from the proper distance, 
which cannot be embedded 
in 3D flat space.
}
\label{fig:3Dplot-s}
\end{center}
\end{figure}

Figs.~\ref{fig:c-s} and \ref{fig:r-s} show the results of $C$ and $r$ 
for $\kappa m_0 = 25$, respectively. 
We can see in Fig.~\ref{fig:c-s}
that $C$ becomes very small around the neck,
and gets even smaller inside the neck of the wormhole, 
but never goes to zero.
For large negative $r_*$,
it has almost the same behavior as the vacuum solution
we described in Sec.~\ref{sec:behindtheneck}
without incompressible fluid ($m_0=0$).

\begin{figure}
\begin{center}
\includegraphics[scale=0.7,bb=0 0 259 170]{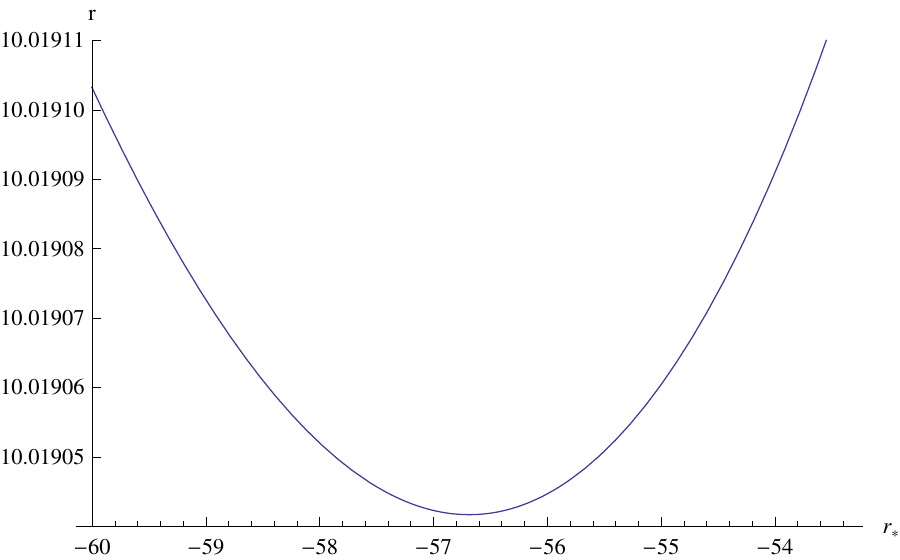}
\hspace{24pt}
\includegraphics[scale=0.7,bb=0 0 259 170]{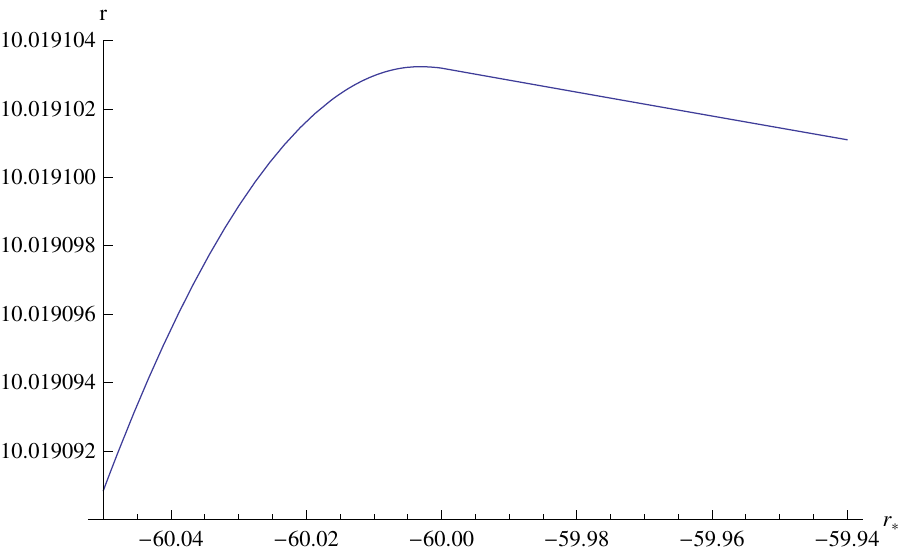}
\caption{\small 
Magnified view of the plot of $r$ vs. $r_*$ around $r_* = a_* \simeq -56.685$ (left) 
and that around 
$r_* \simeq r_{*s}$ (right) of Fig.~\ref{fig:r-s}. 
Note that $r(r_*)$ is in fact a decreasing function 
between $-56.685 < r_* < - 60.003$ ($r$ increases as $r_*$ decreases). 
}
\label{fig:r-ss}
\end{center}
\end{figure}

Fig.~\ref{fig:r-s} shows that the radius $r$ decreases as $r_*$ decreases for $r_*> a_*$,
but $r$ starts to increase with decreasing $r_*$ for $r_* \lesssim a_*$,
where $a_*$ is the tortoise coordinate at the neck.
The radius $r$ starts to decrease again with decreasing $r_*$ from 
the 2nd turning point at $r_* \simeq - 60.003$, 
which is slightly inside the surface $r_{*s}$. 
(Fig.~\ref{fig:r-ss} shows a magnified view of two small regions in Fig.~\ref{fig:r-s}
around $r_* = a_*$ and $r = r_{*s}$.)
In Fig.~\ref{fig:r-s},
we notice that $r$ starts to increase again with decreasing $r_*$ from the 3rd turning point at $r_*\simeq -1097$, 
and never goes to zero but $r\to\infty$ in $r_*\to -\infty$. 
This is also compatible with the vacuum solution described in Sec.~\ref{sec:behindtheneck}. 

The resemblance of the limit $r_*\to - \infty$ of the solution
with the vacuum solution in Sec.~\ref{sec:behindtheneck}
implies that there is a singularity for positive mass in the limit $r_* \to - \infty$.
We deduce that the density $\kappa m_0 = 25$ is too small to have a regular solution
(for $a_0 = 10$ and $r_{*s} = -60$).

\subsubsection{Too large density: $m_0 \gg \hat{m}_0(a_0, r_{*s})$}

\begin{figure}
\begin{center}
\includegraphics[scale=0.7,bb=0 0 259 170]{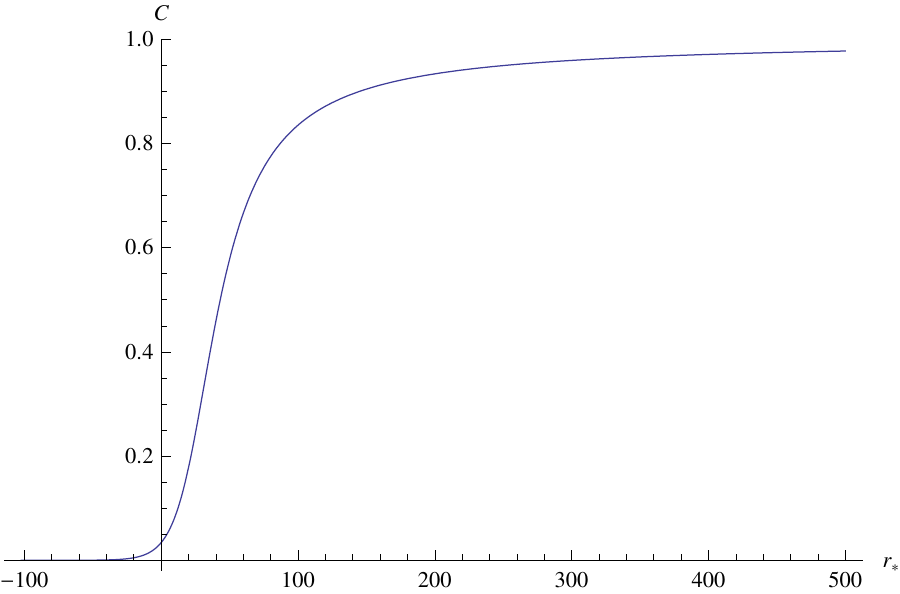}
\hspace{24pt}
\includegraphics[scale=0.7,bb=0 0 259 170]{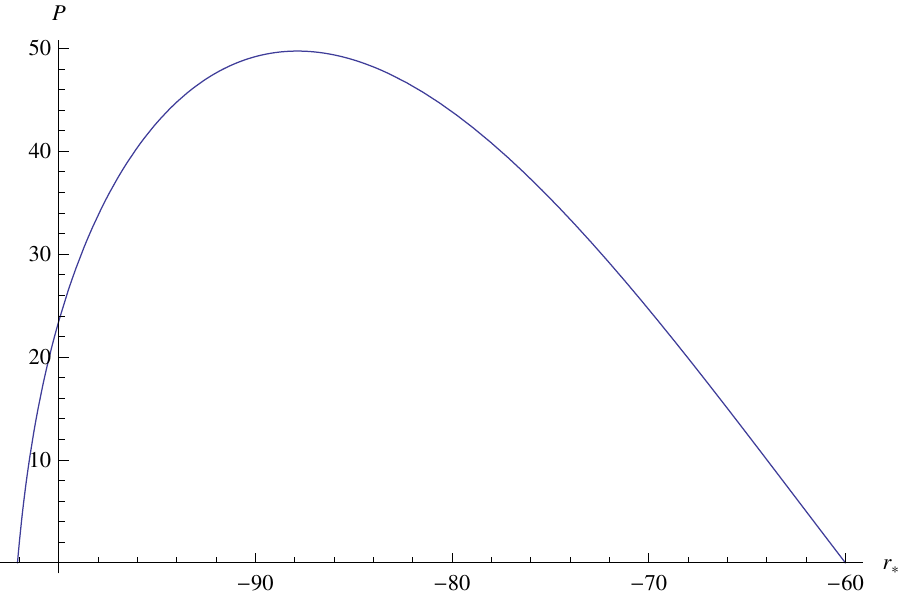}
\caption{\small 
The factor $C$ and pressure $P$
for $\kappa m_0 = 50$, $r_{*s}=-60 < a_*$, $a_0=10$ and $\alpha=0.05$. 
On the left:
The factor $C$ is always positive and non-zero. 
On the right:
The pressure $P$ is zero at both the outer surface $r_*= -60$,
and the inner surface $r_* \simeq -102.086$,
and it is positive in between.
}
\label{fig:c-l}
\end{center}
\end{figure}

\begin{figure}
\begin{center}
\includegraphics[scale=0.7,bb=0 0 259 170]{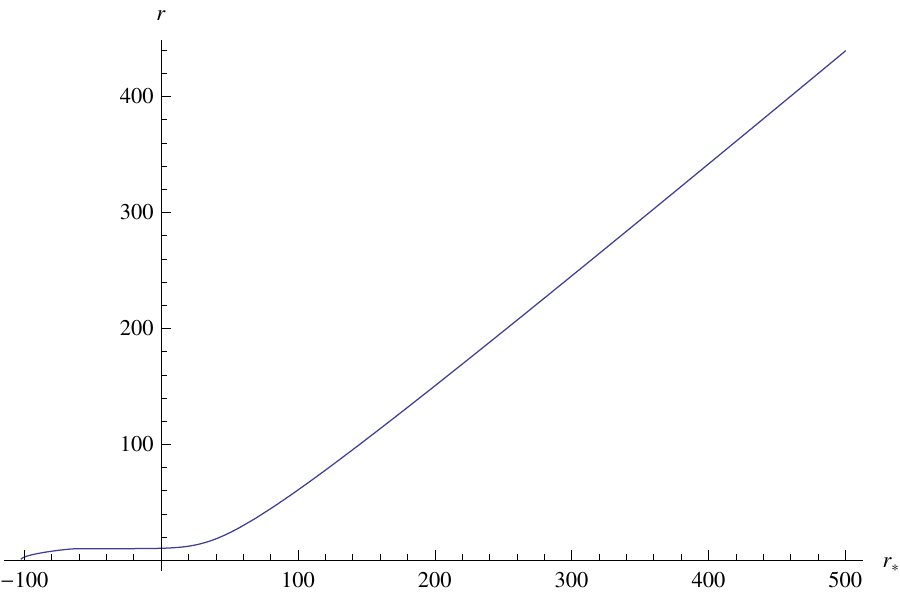}
\hspace{24pt}
\includegraphics[scale=0.7,bb=0 0 259 170]{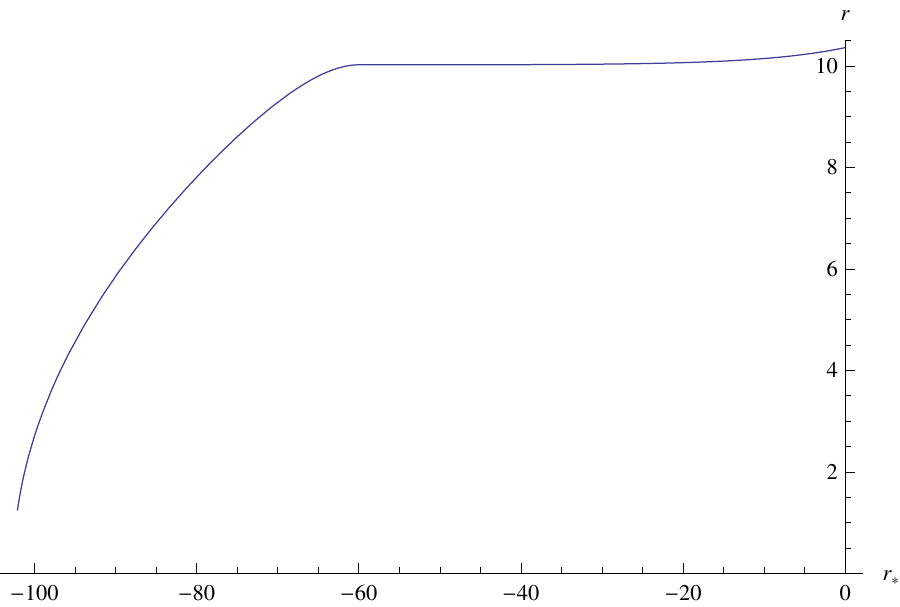}
\caption{\small 
The radius $r$ for $\kappa m_0 = 50$, $r_{*s}=-60 < a_*$, $a_0=10$ and $\alpha=0.05$. 
The plot on the left shows that
the radius $r$ increases as $r_*$ decreases
just inside the neck, 
but starts to decrease from $r_* \simeq - 60.003$.
The plot on the right shows a magnified view of the same diagram of $r$ vs. $r_*$
starting from the inner boundary of the incompressible fluid
at $r_* = r_*^{\tiny inner} \simeq -102.086$ (where $r \simeq 1.25$),
where $P$ vanishes,
up to the point $r_* = 0$.
}
\label{fig:r-l}
\end{center}
\end{figure}

The functions $C(r_*)$ and $r(r_*)$ for a larger density $\kappa m_0=50$ are shown in 
Figs.~\ref{fig:c-l} and \ref{fig:r-l}. 
We start from the asymptotic region without matter
and extend the solution to the internal space of the star with 
its surface located at $r_{*s} =-60$. 
The constant $P_0$ is chosen such that
the pressure $P$ is zero at the surface $r_*= -60$, 
but it turns out that the pressure becomes zero again at $r_* = r_*^{\tiny inner} \simeq -102.086$. 
We stop the calculation at this point since the pressure should be positive, 
but the radius $r$ at this point is still finite: $r^{\tiny inner} \simeq 1.25$.
The physical interpretation of this result 
is that the incompressible fluid is distributed in a shell of finite thickness 
with an inner radius $r_*^{\tiny inner}$ and an outer radius $r_{*s}$.

With $a_0$ and $r_{*s}$ fixed,
a larger density $m_0$ implies a larger inner radius of the star
and a narrower distribution of the incompressible fluid.
For example,
the radius at the inner boundary is $r^{\tiny inner} = 3.64$ for $\kappa m_0 = 100$,
as opposed to $r^{\tiny inner} = 1.25$ for $\kappa m_0 = 50$.

Figs.~\ref{fig:c-l} and \ref{fig:r-l} show the results up to this point at
$r_* = r_{*}^{\tiny inner}$. 
The factor $C$ becomes very small near and inside the neck,
but is still finite as in the previous case. 
Fig.~\ref{fig:r-l} shows that the radius $r$ increases as $r_*$ decreases just inside the neck, 
but starts to decrease from a point $r_* \simeq - 60.003$, 
which is almost on top of but slightly inside the surface of the star
($r_{*s} = - 60$). 
In contrast to the previous case, 
there is no third turning point of $r$,
and $r$ continues to decrease.
Then the pressure becomes zero at $r_* = r_*^{\tiny inner} \simeq -102.086$. 
The second turning point $r_* \simeq - 60.003$ is expected to be larger than 
the previous case,
since the density of the mass is larger.
But the difference is too small to be distinguished. 
The behavior around this point is basically the same as the previous case in Fig.~\ref{fig:r-ss}, 
but the behavior for $r_* < - 60.003$ is a faster decrease in $r$.

If one continues the calculation to smaller values of $r_*$,
assuming that there is no more matter,
there would be a singularity at the origin.
We believe that it is a singularity with negative mass.
The perfect fluid is a shell with finite thickness
in the range $r_*^{\tiny inner} < r_* < r_{*s}$ surrounding the singularity
with vacuum in between.
We explain the stability of the matter shell against gravitational collapse
by the repulsive force from the negative mass at the singularity. 
We conclude that
the density $\kappa m_0=50$ is too large for $a_0 = 10$ and $r_{*s} = -60$.

\subsubsection{Density for regular geometry: $m_0 \sim \hat{m}_0(a_0, r_{*s})$}\label{sssec:num-in-c}

\begin{figure}
\begin{center}
\includegraphics[scale=0.7,bb=0 0 259 170]{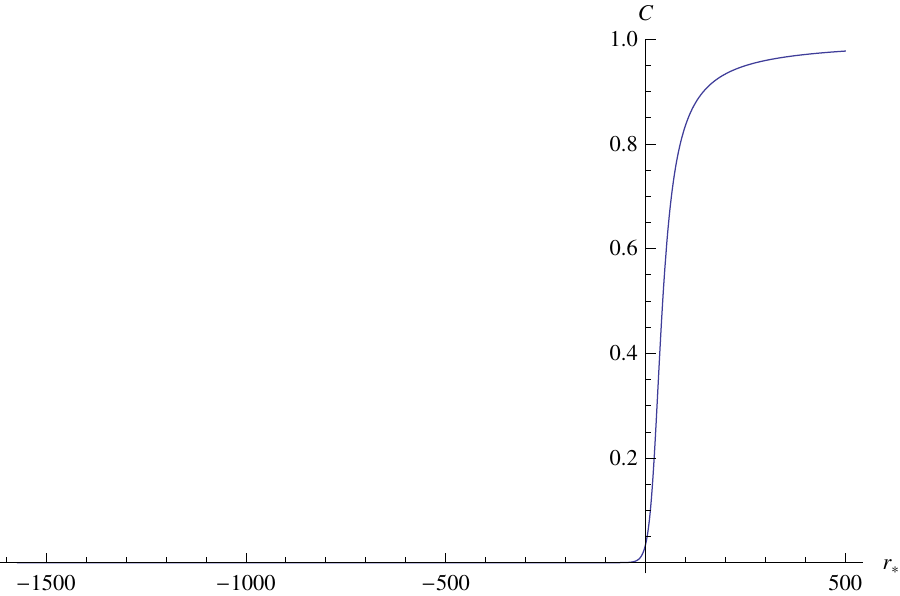}
\hspace{24pt}
\includegraphics[scale=0.7,bb=0 0 259 170]{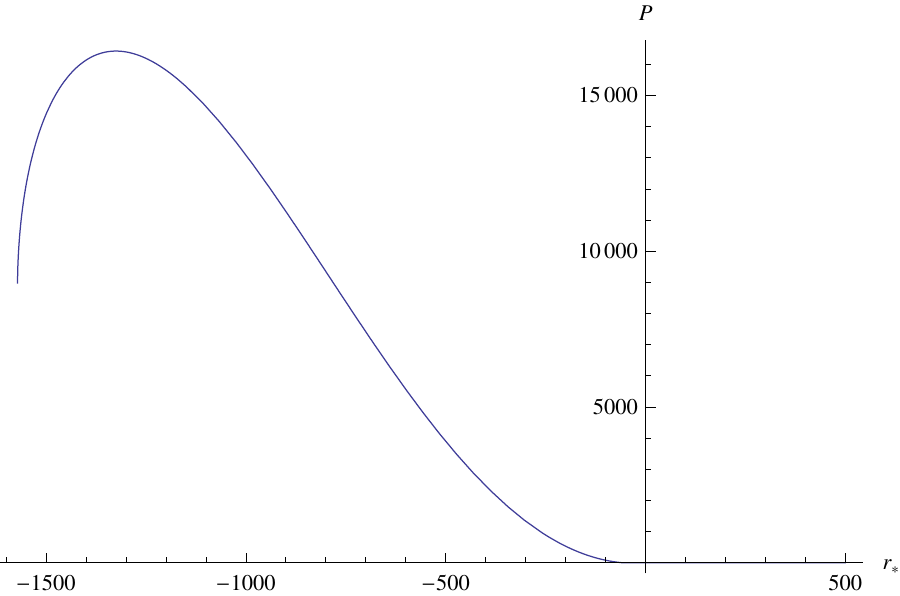}
\caption{\small 
The factor $C$ and the pressure $P$ for 
$\kappa m_0 = 25.2$, $r_{*s}=-60 < a_*$, $a_0=10$ and $\alpha=0.05$. 
The factor $C$ is always non-zero. 
The pressure $P$ is finite everywhere. 
It vanishes at the outer surface of the star
and is positive under the surface.
}
\label{fig:c-c}
\end{center}
\end{figure}

\begin{figure}
\begin{center}
\includegraphics[scale=0.7,bb=0 0 259 170]{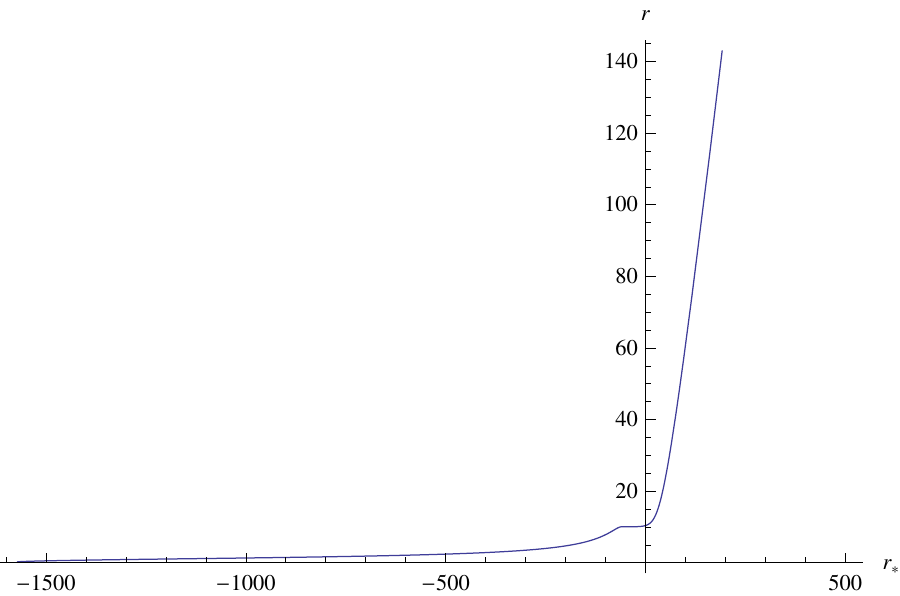}
\hspace{24pt}
\includegraphics[scale=0.7,bb=0 0 259 170]{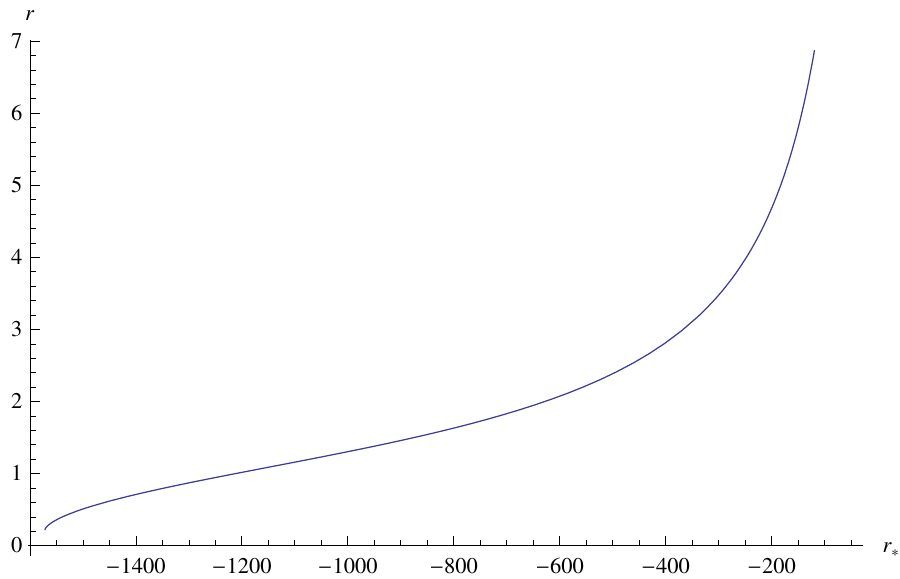}
\caption{\small 
The radius $r$ for $\kappa m_0 = 25.2$, $r_{*s}=-60 < a_*$, $a_0=10$ and $\alpha=0.05$. 
The behavior near the neck is the same as Fig.~\ref{fig:r-ss}. 
The radius $r$ increases as $r_*$ decreases from $r_* = a_* \simeq -56.685$, 
but starts to decrease from $r_* \simeq - 60.003$. 
The instability of the numerical analysis appears
around $r_* \simeq -1571$ where the radius is $r = r_e \simeq 0.22$. 
}
\label{fig:r-c}
\end{center}
\end{figure}

The previous two cases turn out to have too small or too large densities.
There should be an appropriate value 
of the density $m_0 \sim \hat{m}_0$ in between these two cases.
Indeed, we found another type of behavior for the intermediate values of $m_0$.
Figs.~\ref{fig:c-c} and \ref{fig:r-c} show the results for $\kappa m_0 = 25.2$. 
This is a numerical result that fits well with the ideal case of $m_0 = \hat{m}_0$,
so that the geometry is regular everywhere $r \geq 0$,
with finite and positive $P$ inside the surface $r < r_s$.

Regarding the geometry under the neck,
when the density is too small ($m_0 < \hat{m}_0$),
the radius $r$ does not approach to zero but approaches to infinity;
when the density is too large ($m_0 > \hat{m}_0$),
$r$ approaches to zero,
but the pressure $P$ goes to zero first when $r$ is still positive.

Due to numerical instability,
we could not obtain the numerical results with sufficient accuracy
for $r$ smaller than a certain value $r_e$,
which is always found to be of the order of $\sqrt{\alpha}$.
The numerical results are reliable only for $r$ larger than $r_e$.
The idealistic regularity condition that $P$ is finite and positive at $r = 0$ 
for the case $m_0 = \hat{m}_0$ 
is thus replaced by the practical condition that
$P$ is positive and finite at $r = r_e$,
the boundary of numerical analysis.

The instability in our numerical simulation occurs at $r_* \simeq -1571$,
where the radius is $r = r_e \simeq 0.22$. 
(Note that numerically $r_e^2 \simeq \alpha$.)
As we will check later,
the value of $r_e$ depends on $\alpha$
and is smaller for smaller $\alpha$.
In fact, $r_e$ is always approximately $\sqrt{\alpha}$.
Hence, we believe that the numerical instability of the Einstein equation at small $r$
is a reflection of the properties of the vacuum energy.
It merely imposes a bound on the resolution of our analysis around the origin
that does not affect our interpretation 
about the numerical results for the regions of $r \gg r_e$
where the numerical results are reliable.
We will provide more evidence later to support this conjecture. 

Fig.~\ref{fig:c-c} shows that $C$ becomes small around the star 
but is still nonzero as in the previous cases, 
and the pressure $P$ is nonzero
between $r_e$ and the outer surface $r_{*s}$. 
As depicted in Fig.~\ref{fig:r-c},
the behavior of $r$ around the neck resembles Fig.~\ref{fig:r-ss}, again.
The value of $r$ appears to approach to zero
up to an uncertainty of $r_e$.

\begin{figure}
\begin{center}
\includegraphics*[scale=0.09,bb=0 300 1310 2210]{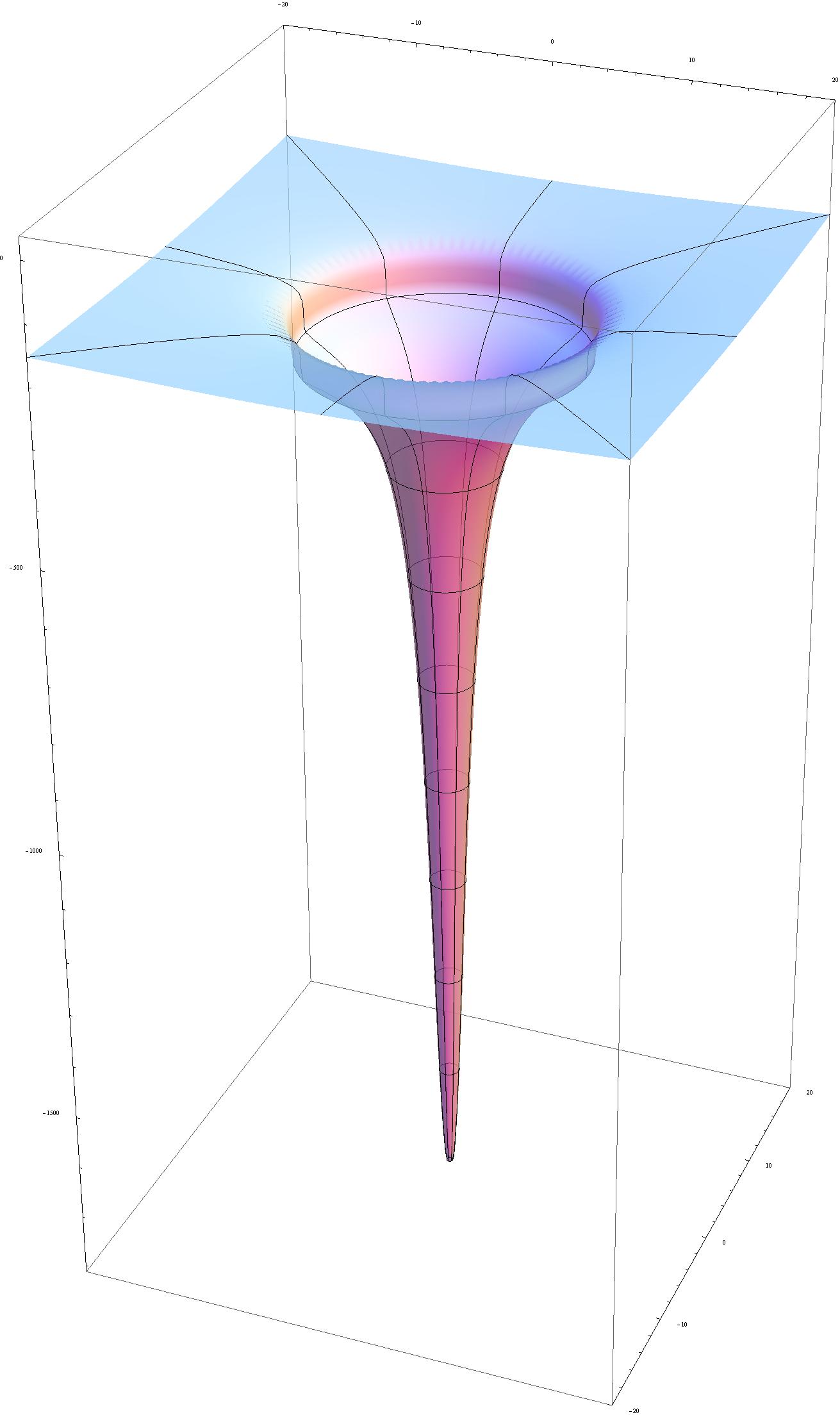}
\caption{\small 
Schematic 3D picture for the case of Fig.\ref{fig:r-c}. 
The radius $r$ corresponds to that in the horizontal direction while  
the vertical axis is chosen such that 
the distance along the 2D surface 
in the picture
is given by the change in the $r_*$ coordinate. 
}
\label{fig:3Dplot-c}
\end{center}
\end{figure}

This result corresponds to the choice of
a approximately appropriate density $m_0 \simeq \hat{m}_0$
for a regular geometry.
It is the transition point between a density too small and a density too large
(or equivalently, a positive or negative mass singularity).
As a numerical simulation,
the determination of the value $\hat{m}_0$ is of course not exact,
and there is a small range of $m_0$ with similar behavior
(e.g., at $\kappa m_0 = 30$). 

We understand its numerical uncertainty as follows.
When $m_0$ is too small ($m_0 < \hat{m}_0$),
there is a third turning point of $r$ under the neck.
However, if the radius at the third turning point is as small as $r_e$, 
we will not be able to distinguish the behavior of the solutions
outside the third turning point from that for the regular solution ($m_0 = \hat{m}_0$). 
If $m_0$ is too large ($m_0 > m_*$),
the perfect fluid is a thick shell with an inner boundary. 
However, if the radius of the inner boundary is as small as $r_e$, 
we cannot distinguish it from that for the regular solution ($m_0 = m_*$).

In this work,
we do not plan to pursue the physics in the tiny region around the origin of scale $\sqrt{\alpha}$,
and we aim to focus on geometry at a larger scale.
As we will see later,
the instability of the numerical calculation in the region of scale $r_e$ around the origin
appears even when the surface of the star is well above the neck
and satisfies the condition \eqref{ClassCond} 
for which the geometry is regular and horizonless in the classical limit. 
It should be clear that
the potential singularity hidden by the numerical instability 
is of a different nature from the singularity at the origin in the classical black hole geometry. 

\begin{figure}
\begin{center}
\includegraphics[scale=0.7,bb=0 0 259 170]{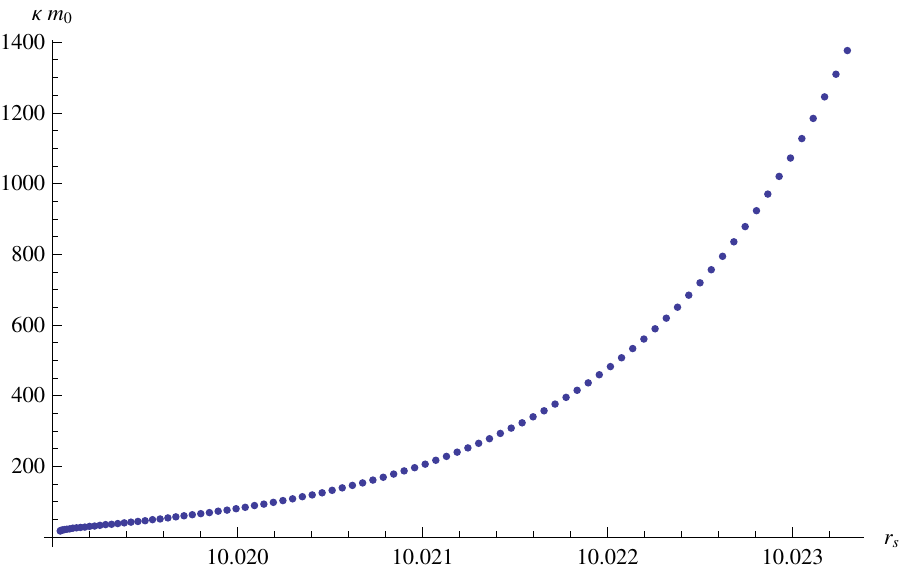}
\hspace{24pt}
\includegraphics[scale=0.7,bb=0 0 259 170]{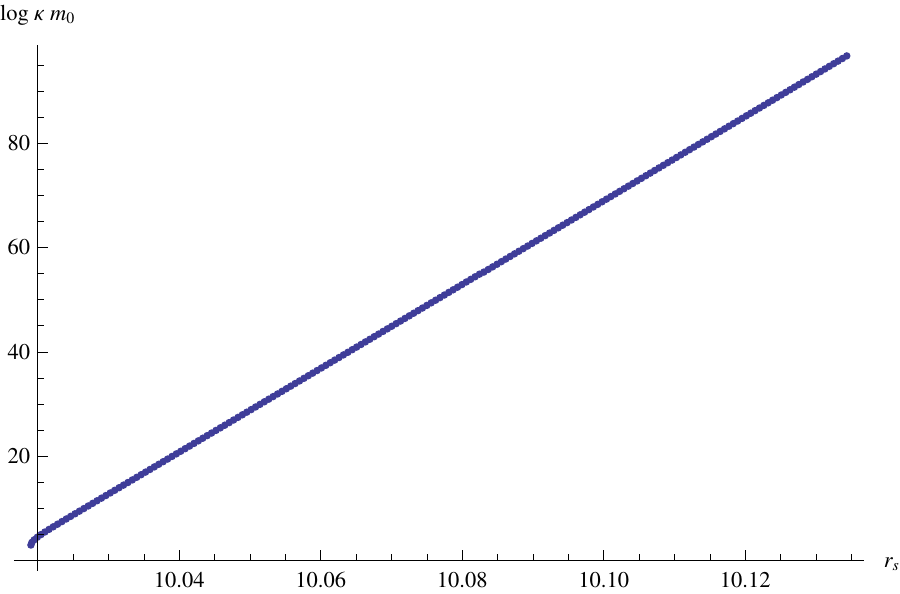}
\caption{\small 
Relation between $\hat{m}_0$ and and $r_{s}$ for $a_0=10$ and $\alpha=0.05$. 
The density $m_0$ increases exponentially as $r_s$ increases.
}
\label{fig:density}
\end{center}
\end{figure}

We now comment on the relation between 
the value of $\hat{m}_0(a_0, r_{*s})$ and $r_{*s}$,
with $a_0$ fixed.
Numerically,
we estimate the value of $\hat{m}_0$ by the smallest value of $m_0$
for which $r$ does not have the third turning point. 
The resulting relation is shown in Fig.~\ref{fig:density}. 
While $r$ increases as $r_*$ decreases under the neck,
$d\ell/dr$ decreases, 
where $\ell$ is the proper distance.
The volume of the star is smaller for larger $r_s$
(with smaller $r_{*s}$)
by a factor of $C^{1/2}$,
and hence larger density $\hat m_0(a_0, r_{*s})$ 
is necessary to keep $a_0$ unchanged. 
Furthermore,
the classical Schwarzschild radius $a_0$ is proportional 
to the mass observed at the infinity,
to which the local mass density $m_0$ contributes 
through the redshift factor $C^{1/2}$. 
With both effects combined,
the total mass $a_0$ is related to the density $m_0$ with a factor of $C$. 
This implies that the density $m_0$ is proportional to $C^{-1}(r_s)$. 
The factor $C$ is exponentially suppressed as $r$ becomes larger,
and hence $\hat m_0(a_0, r_{*s})$ increases exponentially as $r_s$ increases. 

Consistent with this qualitative picture,
Fig.~\ref{fig:density} shows that 
the density $\hat{m}_0$ appears to increase exponentially as $r_s$ increases. 
Since a much large density $\hat{m}_0$ is required for a slightly larger $r_s$,
it is natural to assume that in a generic case
the radius of the surface $r_s$ is only slightly larger than
the radius of the neck $r=a$
unless the density is trans-Planckian.

\begin{figure}
\begin{center}
\includegraphics[scale=0.7,bb=0 0 259 170]{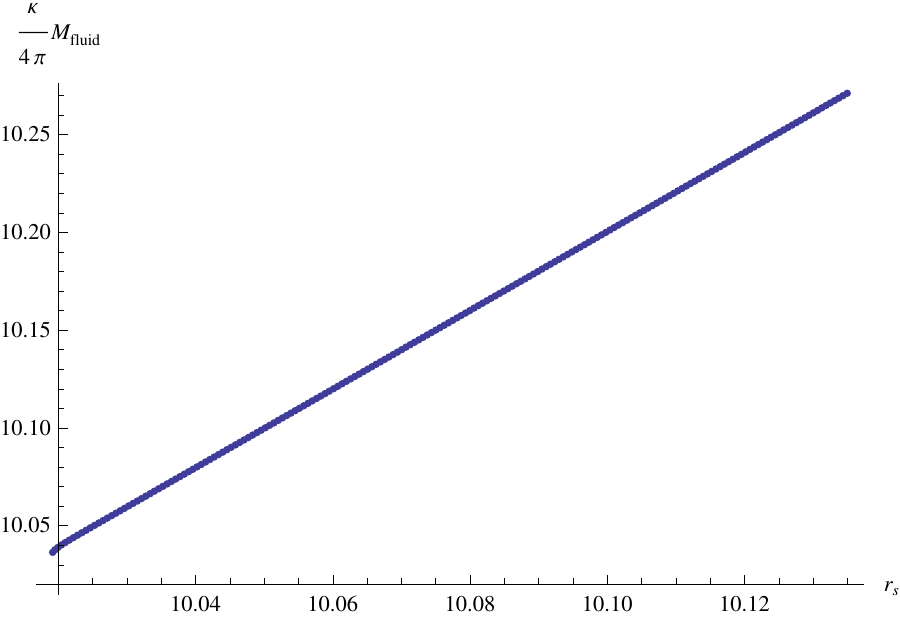}
\hspace{24pt}
\includegraphics[scale=0.7,bb=0 0 259 170]{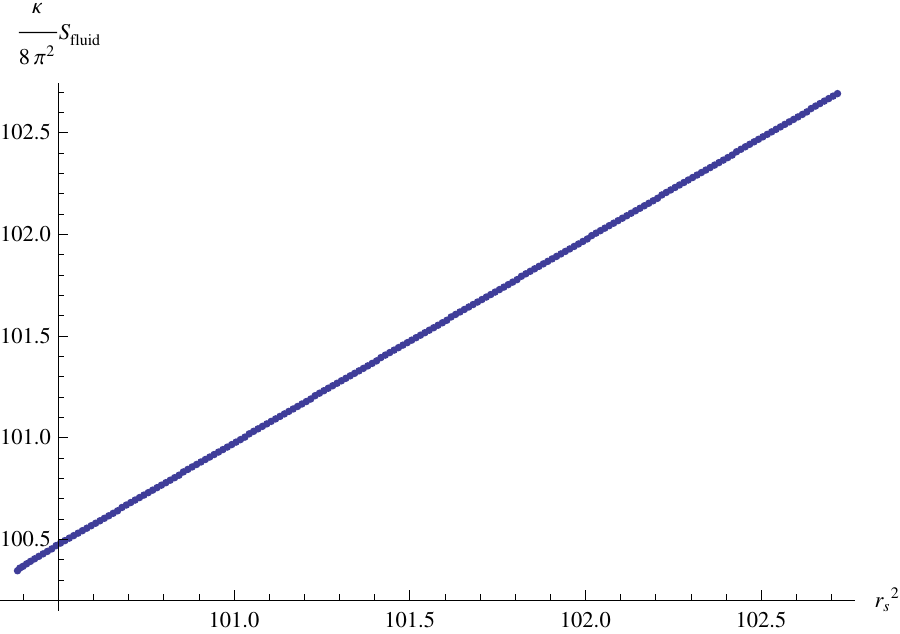}
\caption{\small 
The fluid's contribution to the Komar mass $M_\text{fluid}$ 
and the entropy $S_\text{fluid}$ 
for $a_0=10$ and $\alpha=0.05$.
}
\label{fig:komar-rs}
\end{center}
\end{figure}

The Komar mass of the star is well approximated 
by the density and pressure of the fluid, 
and contribution from the negative vacuum energy is very small. 
For a given classical Schwarzschild radius $a_0$
(a given total mass observed at spatial infinity),
Fig.~\ref{fig:komar-rs} shows 
the contribution of the perfect fluid to the Komar mass:
\begin{equation}
 M_\text{fluid} 
 = - \int d^3 x \sqrt{-g} \left(2 T^0_0 - T^\mu{}_\mu\right) 
 = 4\pi \int dr_*\,r^2 C \left(m_0 + 3 P\right) \ , 
\end{equation}
where the integration is from $r_e$ to $r_s$,
assuming that the contribution from $r<r_e$ is negligible. 
The fluid's contribution to the Komar mass is almost equal to
but slightly larger than the total mass $4\pi a_0/\kappa$,
due to the negative vacuum energy. 

The entropy of the fluid also approximately agrees with 
the Bekenstein-Hawking entropy. 
The entropy density $s$ of the fluid is estimated by 
using the local thermodynamic relation as 
\begin{equation}
 s = \frac{m_0 + P}{T} \ , 
\end{equation}
where $T$ is the local temperature 
which is related to the Hawking temperature $T_H$ by $T = T_H / \sqrt{-g_{tt}}$. 
Here, we assume that the Hawking temperature is simply given by 
that for the classical Schwarzschild black hole, 
and then, the entropy of the fluid is calculated as 
\begin{equation}
 S_\text{fluid} = \int d^3 x \sqrt{h} \ s
 = (4\pi)^2 \int dr_*\,a_0 r^2 C \left(m_0 + P\right) \ , 
\end{equation}
where $h$ is the induced metric on the time-slice. 
The mass $M_\text{fluid}$ and entropy $S_\text{fluid}$ of the fluid 
as functions of $r_s$ for $a_0=10$ and $\alpha=0.05$ is shown in Fig.~\ref{fig:komar-rs}. 
We will show their dependence on $a_0$ later in Fig.~\ref{fig:komar-a0}. 

\subsection{Star above the neck}

We have discussed above the situation when the surface of the star
hides behind the neck of the wormhole-like geometry. 
As we have seen in Section~\ref{sec:Classical}, 
the geometry has the singularity in the classical case 
even if the surface of the star is outside the Schwarzschild radius, $r_s>a_0$. 
Here, we consider the class of solutions with the surface of the star above the neck,
i.e. $r_{*s} > a_{*}$, to see that the singularity in the classical case 
is modified by the negative vacuum energy. 
We shall consider separately two cases:
the case when $r_{*s}$ is slightly larger but very close to $a_{*}$, 
for which the singularity appears in the classical case 
($r < \frac{9}{8} a$),
and the case when $r_{*s}$ is much larger than $a_{*}$, 
for which the geometry is regular in the classical case 
($r > \frac{9}{8} a$).

First we consider the case when $r_{*s}$ is slightly larger but close to $a_{*}$ so that $r < \frac{9}{8} a$. 
As the condition \eqref{CondRadius} is violated,
there is no static regular solution
for the classical Einstein equations with zero vacuum energy.
After taking into account the quantum effects, 
those stars which break this condition \eqref{CondRadius} can also exist. 

There are again three types of solutions.
In these semi-classical solutions,
there is no turning point of $r$ if the density $m_0$
is the same as or larger than the density $\hat{m}_0$
for regular geometry
($m_0 = \hat{m}_0$ or $m_0 > \hat{m}_0$).
On the other hand,
there is one turning point of $r$,
as the case for the vacuum solution
if the density $m_0$ is smaller than $\hat{m}_0$
($m_0 < \hat{m}_0$). 

The turning point for this case $m_0 < \hat{m}_0$ is slightly smaller than 
the radius $r = a$ of the neck for the case 
when all the matter resides behind the neck.
This is because the size of the turning point is determined by
the total mass inside the turning point,
and the latter is smaller 
as some of the matter resides outside the turning point. 

\begin{figure}
\begin{center}
\includegraphics[scale=0.7,bb=0 0 259 170]{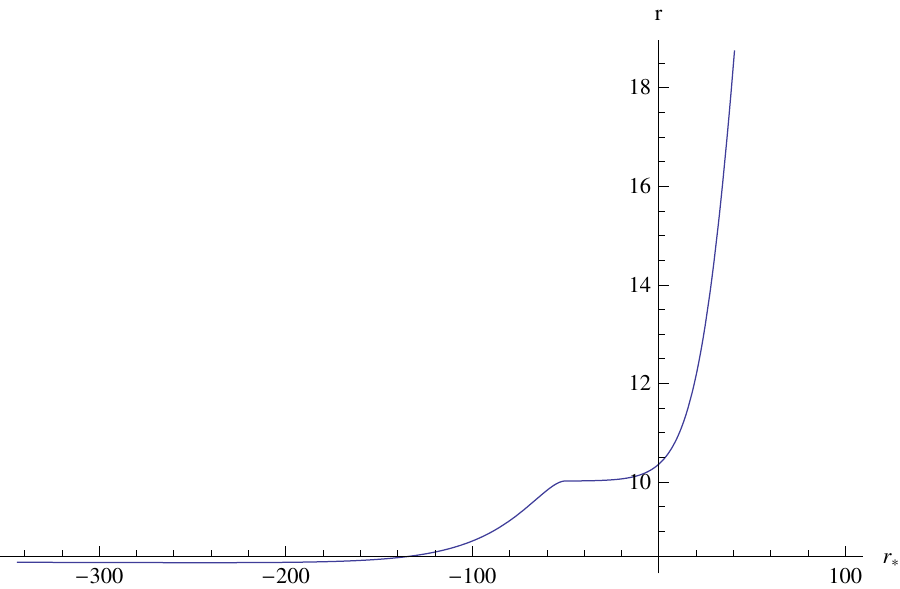}
\hspace{24pt}
\includegraphics[scale=0.7,bb=0 0 259 170]{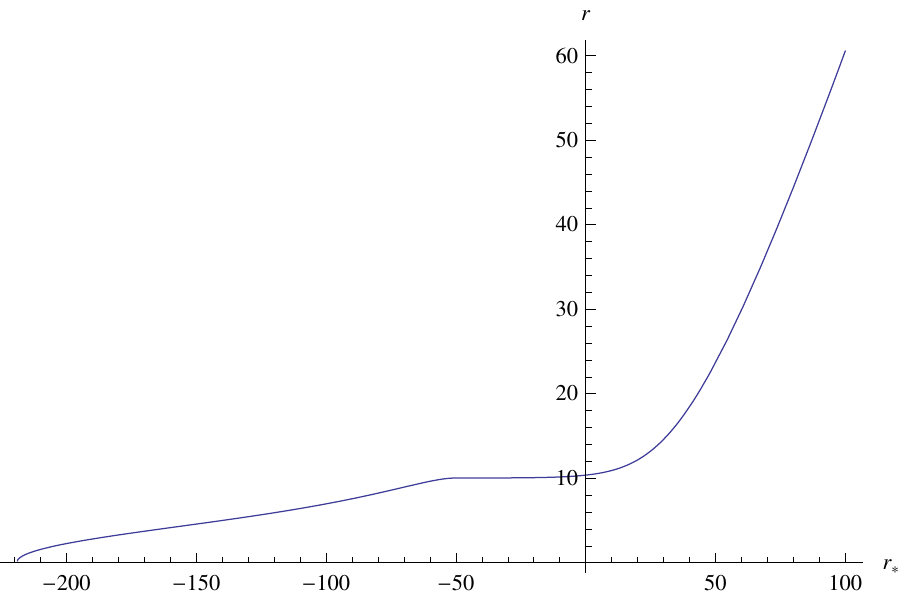}
\caption{\small 
Behavior of $r$ for the surface $r_s$ close to but outside the neck at $r=a$. 
Here we take $r_{*s} = -50 > a_* \simeq -56.685$, $a_0=10$ and $\alpha=0.05$.
For $\kappa m_0 = 5$ (left), the turning point of $r$ is around $r_* = -243$, where $r = 8.36$. 
For $\kappa m_0 = 10$ (right), there are no turning point of $r$ 
and $r$ approaches to $r_e$ scale at finite $r_*$. 
}
\label{fig:r-o}
\end{center}
\end{figure}

For example, for the surface radius at $r_{*s} = -50$, where $r_{s} = 10.0194$, 
the radius for regular geometry is around $\kappa m = \hat{m}_0 \simeq 10$, 
and $\kappa m_0 = 5$ is too small while $\kappa m_0 = 20$ is too large. 
For $\kappa m_0 = 5$, the turning point is around $r_* = -243$, where $r = 8.36$. 
Fig.~\ref{fig:r-o} shows the behavior of $r$ for $\kappa m_0 = 5$ and $\kappa m_0 = 10$. 
The behavior of $r$ for $\kappa m_0 = 20$ is almost the same as that for $\kappa m_0 = 10$,
but the pressure $P$ vanishes where $r$ is much larger than $r_e$.

Let us now consider the case when the surface of the star 
is well above the neck, i.e. $r_{*s} \gg a_{*}$
so that the condition $r > \frac{9}{8} a$ \eqref{CondRadius} is satisfied,
and the geometry is well defined without divergence even in the classical limit. 

\begin{figure}
\begin{center}
\includegraphics[scale=0.7,bb=0 0 259 170]{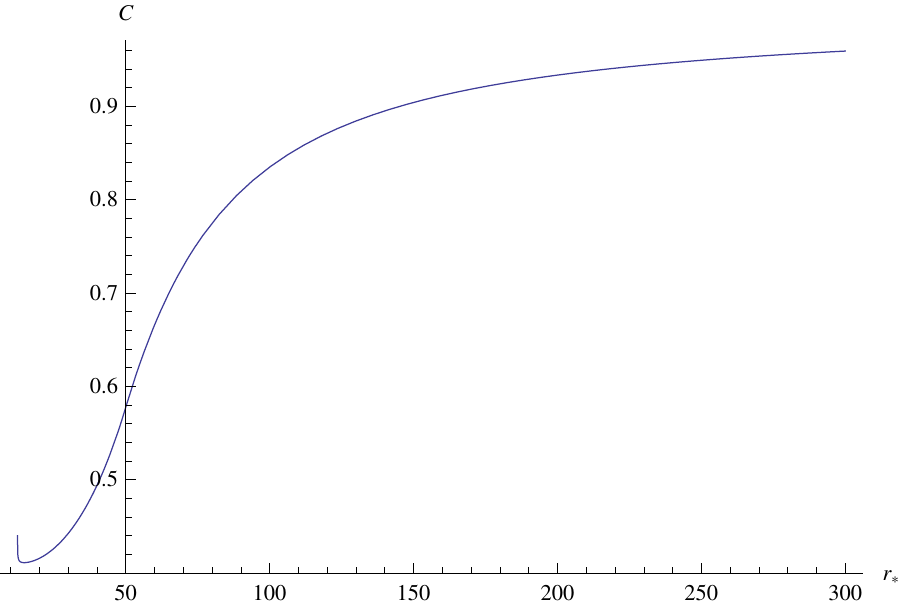}
\hspace{24pt}
\includegraphics[scale=0.7,bb=0 0 259 170]{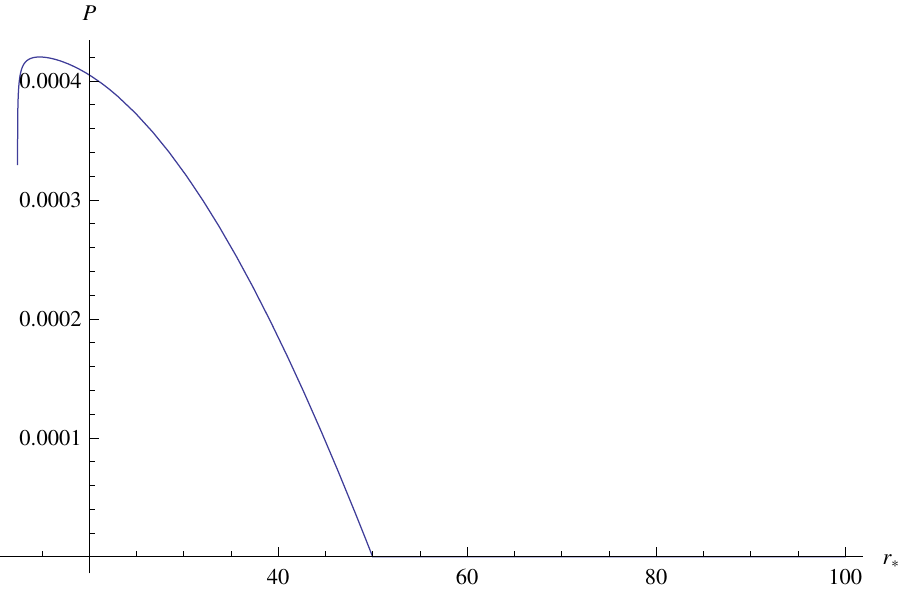}
\caption{\small 
The factor $C$ (left) and the pressure $P$ (right) 
for $\kappa m_0 \simeq \kappa m_* \simeq 0.00227$ with the surface radius
$r_{*s} = 50$ where $r_s \simeq 23.65$. 
}
\label{fig:c-n}
\end{center}
\end{figure}

\begin{figure}
\begin{center}
\includegraphics[scale=0.7,bb=0 0 259 170]{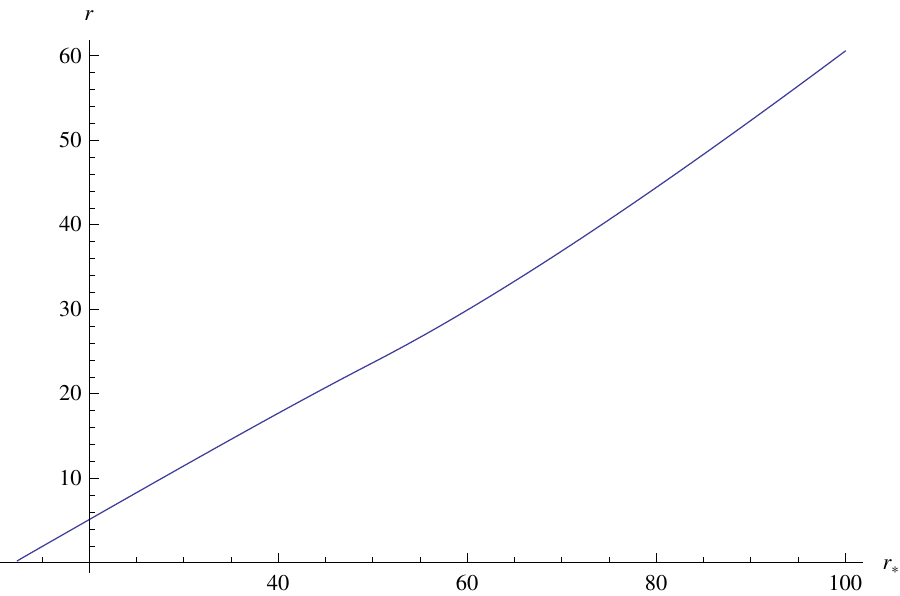}
\hspace{24pt}
\includegraphics[scale=0.7,bb=0 0 259 170]{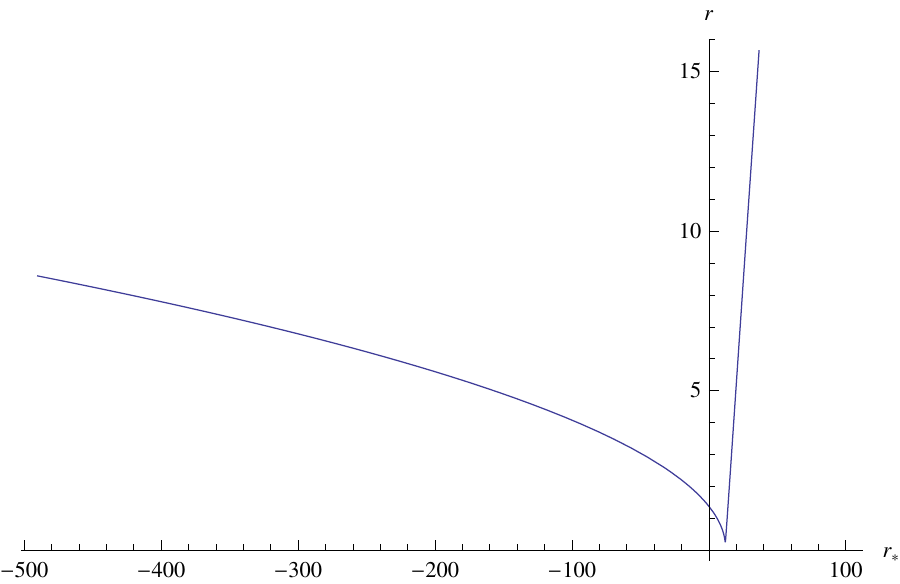}
\caption{\small 
The radius $r$ for for $\kappa m_0 \simeq \kappa \hat{m}_0 \simeq 0.00227$ (left)
and that for $\kappa m_0 = 0.00226$ (right). 
For $\kappa m_0 = 0.00227$, it approaches to zero but 
the numerical difficulty arises around $r \simeq 0.22$, 
while the turning point of $r$ for $\kappa m_0 = 0.00226$ 
is slightly outside of this point: $r \simeq 0.24$. 
}
\label{fig:r-n}
\end{center}
\end{figure}

There are again 3 types of solutions. 
If the density $m_0$ is smaller than $\hat{m}_0$,
there is a turning point of $r$. 
For example, for the surface at $r_{*s} = 50$, with $r_s \simeq 23.65$, 
the density $\hat{m}_0$ for regular geometry is approximately
$\kappa m_0 \simeq \kappa \hat{m}_0 \simeq 0.00227$,
and $\kappa m_0 = 0.00226$ is too small while $\kappa m_0 = 0.00228$ is too large. 
Notice that they approximately satisfy the classical relation between 
the total mass and the density, 
\begin{equation}
 \frac{4\pi}{3} \frac{\kappa}{8\pi} \hat m_0 r_s^3 \simeq 5.0046 \simeq \frac{a_0}{2} \ . 
\end{equation}
Fig.~\ref{fig:c-n} is a plot of $C$ and $P$ for $\kappa m_0 = 0.00227$. 
The radius $r$ for $\kappa m_0 = 0.00227$ and that for $\kappa m_0 = 0.00226$ 
are depicted in Fig.~\ref{fig:r-n}. 
The limit of the numerical calculation due to the numerical instability
mentioned above
is around $r_e \simeq 0.22$ for $\kappa m_0 = 0.00227$, 
while the turning point of $r$ for $\kappa m_0 = 0.00226$ 
is slightly outside the star, at $r \simeq 0.24$. 
Notice that,
since the numerical instability at scale $r_e$ around the origin arises even in this case, 
which is classically not a black hole but an ordinary star, 
this technical problem is not related to the singularity of black holes.

\subsection{Surface on the neck}

In order to see how solutions depend on $a_0$ and $\alpha$, 
we put the surface of the star at the neck of the wormhole-like geometry,
i.e. $r_s = a$. 
We focus on the case with the density $m_0 = \hat{m}_0$ for regular geometry.
If the vacuum energy is not taken into account, 
the density $m_0$ is related to the total mass by $M = \frac{4\pi}{3} m_0 r_s^3$. 
If the surface is outside the wormhole and the surface radius $r_s$ is sufficiently large, 
the quantum corrections are very small and this relation holds approximately. 
In the classical case, the surface radius $r_s$ should satisfy 
$r_s > \frac{9}{8}a_0$ for regularity. 
However, by taking the effects of the vacuum negative energy, 
the geometry is still regular even for $r_s = a$. 
If the continuation of the relation $M = \frac{4\pi}{3} m_0 r_s^3$ to $r_s = a$ 
were approximately correct, 
the density would be related to $a_0$ by $m_0 \sim M r_s^{-3} \sim r_s^{-2} \sim a_0^{-2}$, 
and the density could be arbitrarily small by taking sufficiently large radius. 

\begin{figure}
\begin{center}
\includegraphics[scale=0.7,bb=0 0 259 170]{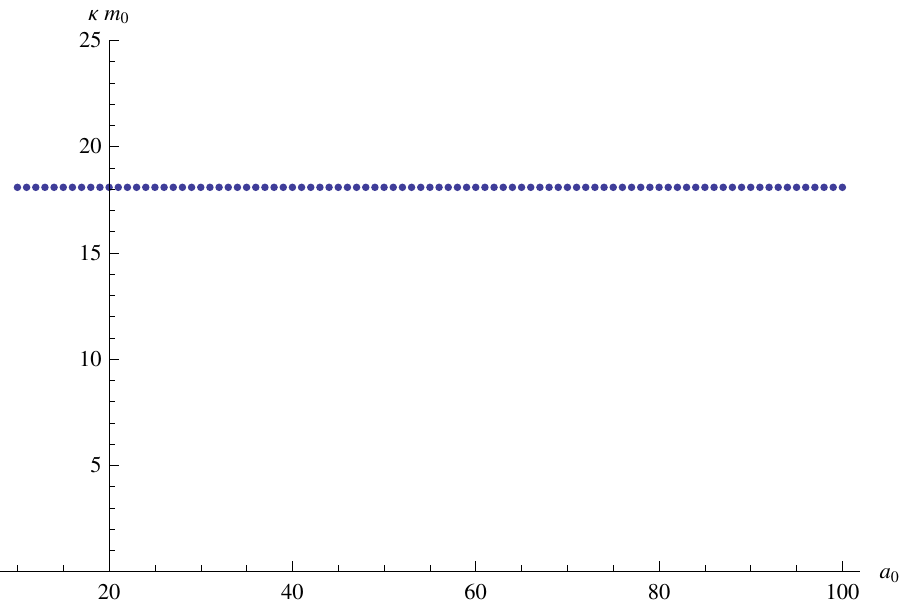}
\caption{\small 
The dependence of $\hat{m}_0$ 
on $a_0$ for $\alpha = 0.05$. 
The surface $r_{s*}$ is located on the neck, $r_{*s} = a_*$. 
The plot shows that $\hat{m}_0$ is almost independent of $a_0$. 
}
\label{fig:m0-a0}
\end{center}
\end{figure}

Let us see what happens when the quantum effect is taken into account. 
The relation between $\hat{m}_0$ and $a_0$ for fixed $\alpha$ is shown in Fig.~\ref{fig:m0-a0}. 
Here we take $\alpha = 0.05$ as we did in previous examples. 
The result shows that the density $\hat{m}_0$ is almost 
independent of the classical Schwarzschild radius $a_0$. 
As in previous examples,
numerical calculation only allows us to fix the value of $\hat{m}_0$ within a finite range, 
for example,
$18.1 \lesssim \kappa \hat{m}_0 \lesssim 22.9$ for $a_0=10$.  
The density $\hat{m}_0$ shown in Fig.~\ref{fig:m0-a0} is around the minimum of this range. 
While the minimum of the range may not properly reflect 
the $a_0$-dependence of the exact value of $\hat{m}_0$,
it is clear from Fig.~\ref{fig:m0-a0} that $\hat{m}_0$ is insensitive to changes in $a_0$ 
even though it appears to slightly increase as $a_0$ increases.

\begin{figure}
\begin{center}
\includegraphics[scale=0.7,bb=0 0 259 170]{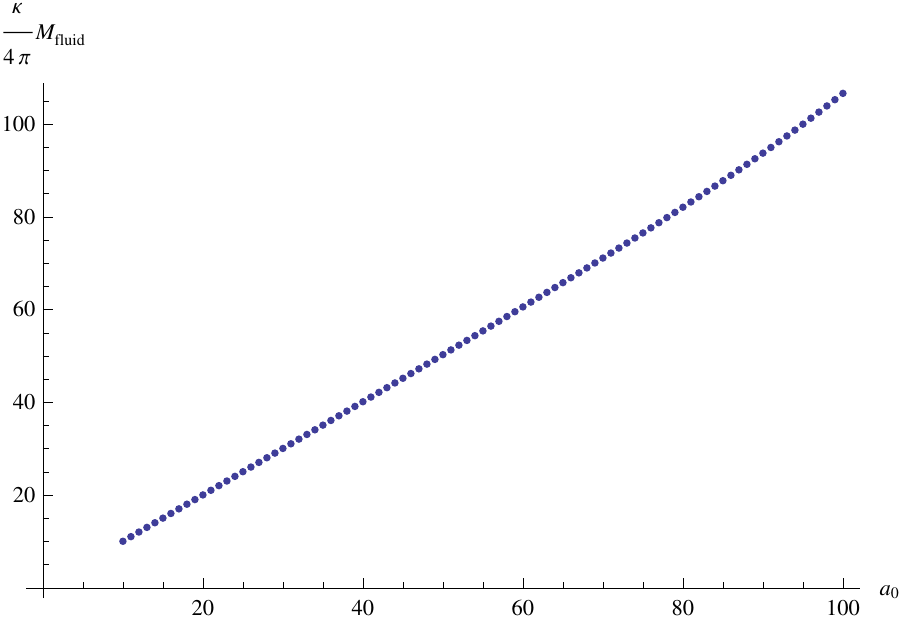}
\hspace{24pt}
\includegraphics[scale=0.7,bb=0 0 259 170]{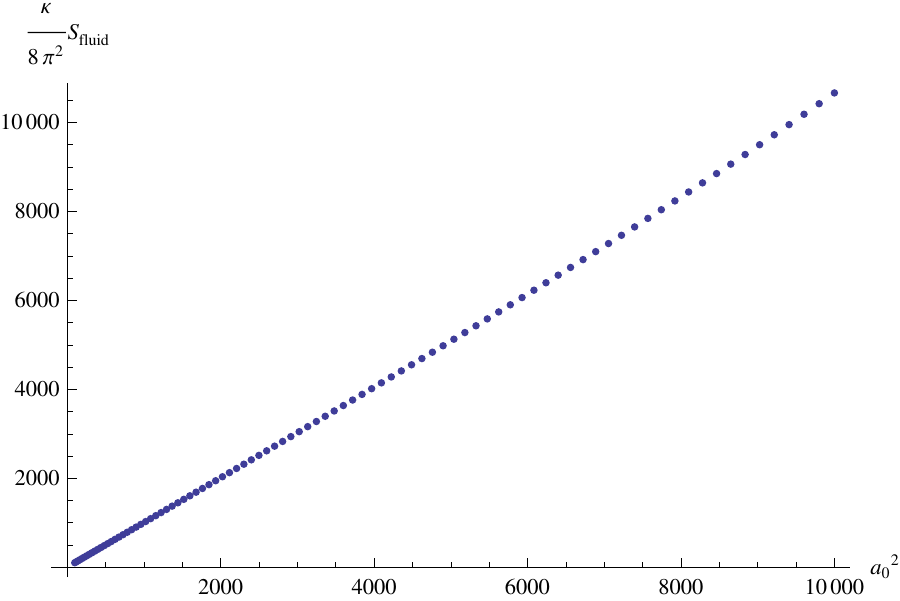}
\caption{\small 
The dependence of total mass $M_\text{fluid}$ and the entropy $S_\text{fluid}$ of the fluid on $a_0$ for $\alpha = 0.05$. 
The surface $r_{s*}$ is located on the neck, $r_{*s} = a_*$.  
}
\label{fig:komar-a0}
\end{center}
\end{figure}

Although the density $\hat m_0$ is almost independent of the classical Schwarzschild radius $a_0$, 
the contributions to the Komar mass $M_\text{fluid}$ and the entropy $S_\text{fluid}$ 
from the fluid alone (without those from the vacuum state) 
well reproduce the classical Schwarzschild radius and 
the Bekenstein Hawking entropy of the black hole. 
The relation of the Komar mass $M_\text{fluid}$ and the entropy $S_\text{fluid}$ 
with the Schwarzschild radius $a_0$ is shown in Fig.~\ref{fig:komar-a0}. 
It should be noted that the classical Schwarzschild radius approximately 
equals to the surface radius since $a \sim a_0$ and we took $r_s = a$. 
Clearly the classical contribution to energy dominates over its quantum counterpart.

\begin{figure}
\begin{center}
\includegraphics[scale=0.7,bb=0 0 259 170]{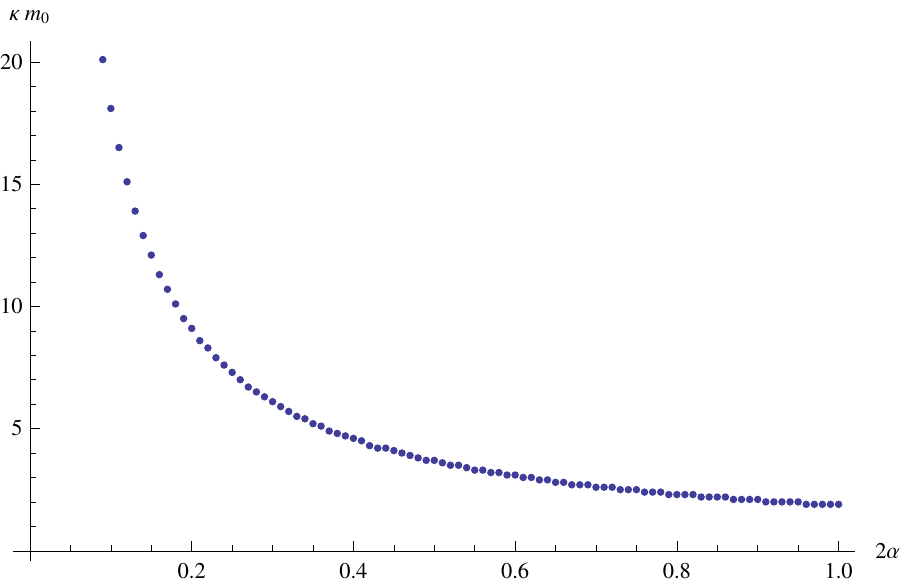}
\hspace{24pt}
\includegraphics[scale=0.7,bb=0 0 259 170]{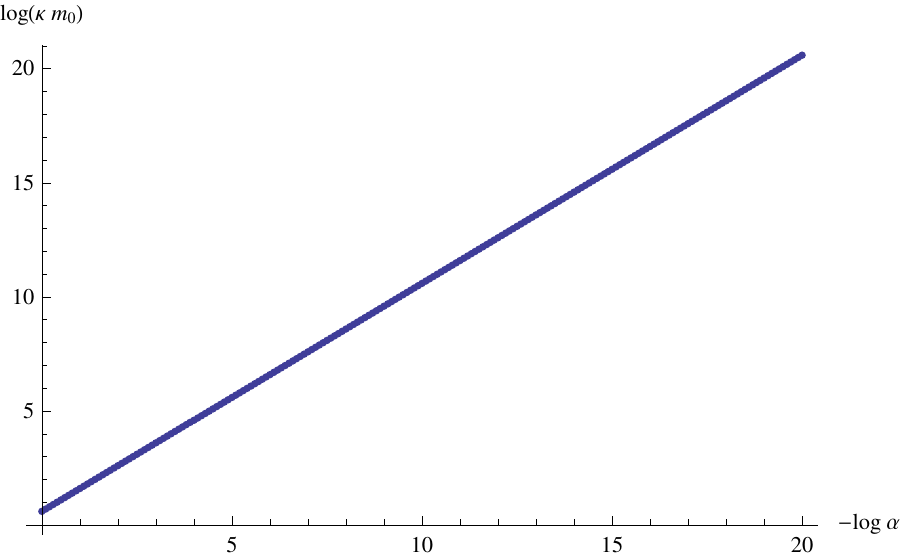}
\caption{\small 
Relation between $m_0$ and $\alpha$ for $a_0 = 10$. 
The surface of the star $r_{*s}$ is chosen to be on the neck ($r(r_{*s}) = a$),
but depends on $\alpha$. 
The density $m_0$ behaves as $m_0 \propto \kappa^{-1} \alpha^{-1}$. 
}
\label{fig:m0-alpha}
\end{center}
\end{figure}

The relation between $\hat{m}_0$ and $\alpha$ 
for fixed $a_0$ is depicted in Fig.~\ref{fig:m0-alpha}. 
Here, we take $a_0= 10$. 
Clearly the density $\hat{m}_0$ and $\alpha$ 
are related to each other as 
\begin{equation}
 \hat{m}_0 \sim \frac{\alpha^{-1}}{\kappa} \ . 
 \label{mak}
\end{equation}
In terms of the Planck length $\ell_p$, 
the density behaves as $\hat{m}_0 \sim N^{-1} \ell_p^{-4}$. 
We assume in this paper that $N$ is very large
so that the density is much lower than the Planck scale.

It should be noted that in principle all fields in the theory 
should contribute to an effective value of $N$
through their contribution to the vacuum energy,
including those which are not present in the classical matter of the star.
Hence the number of the fields in the theory 
could be as large as $\mathcal O(10^2)$, 
in the standard model, for example.
In a hypothetical UV-complete theory in which
the number of fields is arbitrarily large,
$\sqrt{\alpha}$ defines a characteristic length scale 
that is arbitrarily larger than the Planck length.
This would allow us to avoid Planck-scale physics
in our discussions.

\begin{figure}
\begin{center}
\includegraphics[scale=0.7,bb=0 0 259 170]{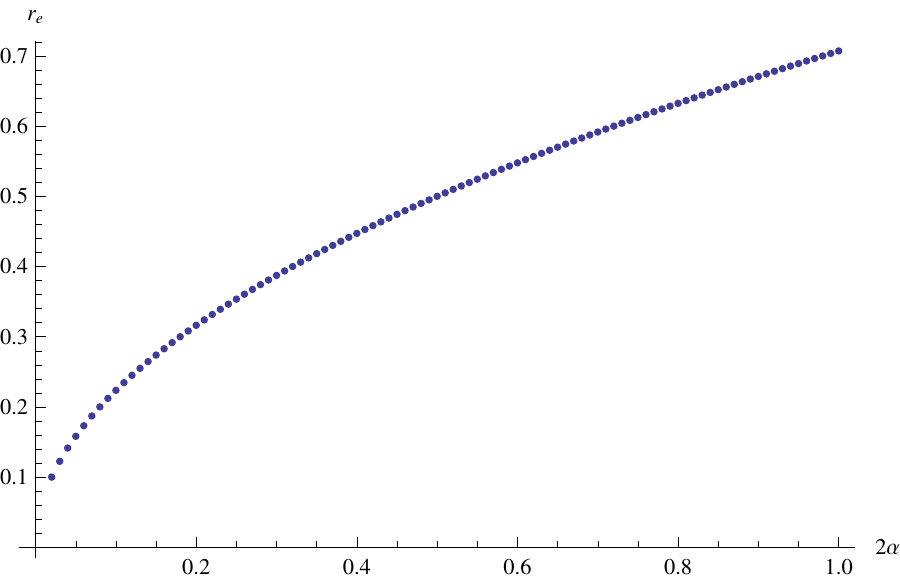}
\hspace{24pt}
\includegraphics[scale=0.7,bb=0 0 259 170]{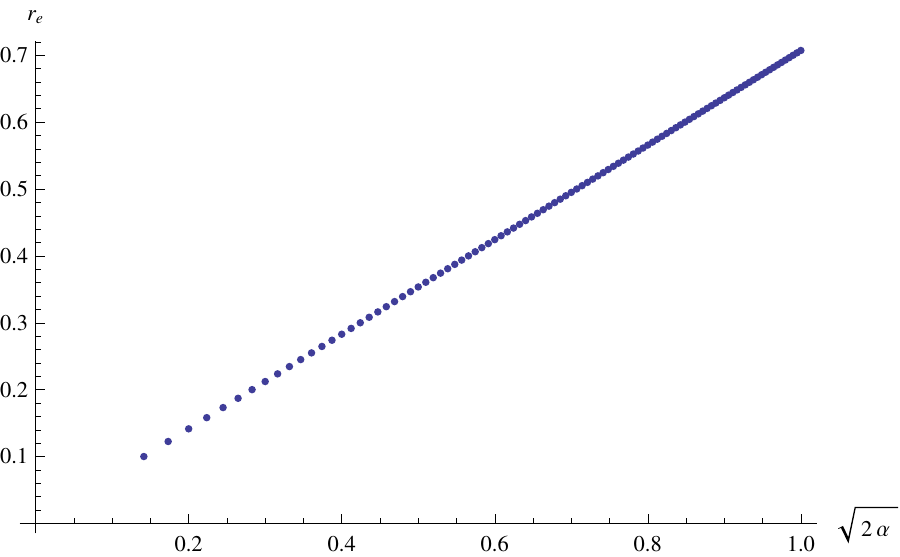}
\caption{\small 
The radius $r = r_e$ below which the numerical calculation is unreliable
due to numerical instability for $a_0 = 10$ but different values of $\alpha$.
The surface of the star $r_{*s}$ is defined by the relation $r(r_{*s}) = a$. 
The radius $r_e$ behaves as $r_e \simeq \alpha^{1/2}$.
}
\label{fig:rmin}
\end{center}
\end{figure}

Finally, Fig.~\ref{fig:rmin} shows 
the radius where we face the technical difficulty, $r = r_e$. 
The radius $r_e$ is, in fact, proportional to $\sqrt{\alpha}$,
and hence the problem is only in a small region around the origin.

\section{Conclusion and discussions}\label{sec:conclusion}

\subsection{Summary}

Solving the semi-classical Einstein equation,
we studied the geometry of a static, spherically symmetric configuration 
with the back reaction of vacuum energy taken into consideration.
The vacuum energy is given by a toy model
that is often used in the study of black holes in the literature.
For the Boulware vacuum
(no incoming or outgoing energy flux at the spatial infinity),
the vacuum has negative energy which diverges at the horizon in the perturbative calculation
\cite{Davies:1976ei}.
For this reason it is conventionally assumed that the Boulware vacuum is not physical 
unless the radius of the star is much larger than the Schwarzschild radius, 
but it was shown in \cite{Ho:2017joh} that the divergence of the negative energy 
is modified by taking the back reaction to the negative energy to the geometry into account. 
In the nonperturbative solution of the semi-classical Einstein equation, 
the vacuum has negative energy without divergence for the Boulware vacuum, 
and the black-hole geometry has no event horizon. 
We emphasize that the expectation value of the energy-momentum tensor 
is obtained by using the exactly same setup as Ref.\cite{Davies:1976ei}, 
in the sense that it is calculated by using 
the same conservation equation and the initial condition. 
The differences are only that we are considering the static configuration of a star 
while a gravitational collapse is considered in Ref.\cite{Davies:1976ei}, 
and that the back reaction from the vacuum energy-momentum tensor 
is taken into account in this paper. 

Around the classical Schwarzschild radius $r=a_0$, 
the geometry has
a local minimum of the radius.
The local minimal radius $r = a$ is called the quantum Schwarzschild radius.
It differs from the classical Schwarzschild radius $a_0$ by 
$a - a_0 \simeq {\cal O}(\alpha/a_0)$, 
corresponding to a proper distance of order ${\cal O}(\sqrt{\alpha})$. 
This difference is associated with 
the negative energy outside the quantum Schwarzschild radius, 
which is only of order $\mathcal O(a_0^{-1})$. 

Assuming a large number $N$ of fields contributing to the vacuum energy,
the quantity $\alpha = \kappa N/24\pi$
defines a characteristic length scale $\sqrt{\alpha}$ of vacuum geometry
that is much longer than the Planck length $\ell_p = \sqrt{\kappa}$.
This is hence a theoretical model in which the quantum gravitational effect
is suppressed while the quantum effect in low energy effective theories
plays an important role.

A star with its surface below the neck at $r = a$ would appear
very much like a black hole to a distant observer
due to the huge blue-shift factor of order ${\cal O}(a/\sqrt{\alpha})$ at the neck.
The space behind the neck has an even larger blue-shift factor,
but it is still causally connected with the space outside,
in the absence of horizon.
Furthermore,
if the surface of the star is below the neck,
its proper distance from the neck is at most of order ${\cal O}(\sqrt{\alpha})$,
This is a result of the peculiar property of the geometry in vacuum
described in Sec.\ref{sec:behindtheneck},
and is thus independent of the matter content of the star.

The mass density of the vacuum around the neck is only of order ${\cal O}(1/a^2\ell_p^2)$.
It is much smaller than the mass density of the incompressible fluid,
which is of order ${\cal O}(1/\kappa\alpha)$ or larger,
if the surface of the star is around the neck or lower.
While the total energy is dominated by the matter,
the vacuum energy plays an important role in 
modifying the geometry so that
the surface of the star is always close to the Schwarzschild radius.

We studied thin shells and incompressible fluid
as simple models of black-hole-like objects.
Due to the effect of the vacuum energy,
classically forbidden configurations are regularized.
There are static configurations with smooth distribution of energy and pressure
which are regular everywhere
(up to an uncertainty within a region of scale $\sqrt{\alpha}$ around the origin).

In the classical case,
it is well known that the pressure diverges for incompressible fluid 
if the surface radius is smaller than $\frac{9}{8}$ of the classical Schwarzschild radius,
or equivalently, if the density $\kappa m_0$ is larger than $\frac{8}{3 r_s^2}$.
Therefore,
for a given incompressible fluid (with given mass density $m_0$),
the classical theory predicts its collapse into a black hole
for sufficiently large radius $r_s$,
even if $r_s$ is initially larger than $a_0$.

By turning on the vacuum energy of a quantum field,
the pressure no longer diverges. 
There is no bound on the radius or the density.
This may be surprising to the reader
because the vacuum energy can be extremely small 
if the radius $r_s$ is not very close to $a_0$.
Yet a small correction to the energy-momentum tensor 
is in fact capable of significant modifications.
We have given an analytic proof of the regularity of the pressure 
in a similar fashion as our proof for the absence of horizon in the vacuum solution.

\subsection{Comments}

One may wonder how a tiny vacuum energy can have the significant effect
on the geometry as described in this paper.
Note that the Hawking radiation is also extremely weak
as part of the vacuum energy-momentum tensor,
but it can lead to the complete evaporation of an arbitrarily large black hole,
making crucial difference to the global causal structure of space-time.
Note also that any radiation,
however weak it is,
can appear arbitrarily strong 
for an observer moving towards the source close to the speed of light.
Whether an energy-momentum tensor is strong or weak
depends on the specific problem we focus on.

While the spacetime outside the Schwarzschild radius is
very well approximated by the Schwarzschild solution,
the modification of the geometry around and below the neck
seems dramatic at first sight.
For instance,
the horizon disappears completely due to the vacuum energy.
However,
the event horizon is defined by the condition that
anything inside the horizon takes an infinite amount of time to get out.
If the geometry is deformed such that 
anything inside the horizon takes an extremely long time
(say, $10^{100}$ times the age of the universe) to come out,
even though the event horizon is completely removed,
it can still be viewed a small deformation for physicists.

For the model of vacuum energy we have considered,
the geometry is modified such that
the surface of the star can never be 
over a few Planck lengths behind the Schwarzschild radius,
where the horizon is replaced by a turning point in the radius.
This feature is reminiscent of the fuzzball scenario \cite{FuzzBall}
and other proposals \cite{Park:2017dib} 
in which quantum effects modify the near-horizon geometry.
This model serves as a new class of order-1 corrections to the horizon geometry,
which is needed in the dynamical process of gravitational collapse
for unitarity \cite{Mathur:2009hf}. 
The large energy density and pressure in matter when 
the surface of the star is close to the Schwarzschild radius 
warrant us to invoke a high energy theory to describe the interaction 
between Hawking radiation and the matter in the star. 
This breakdown of the niceness conditions \cite{Mathur:2009hf} is 
crucial for a resolution of the information loss paradox.

\subsection{Outlook}

It is not very clear whether this model defined by 2D massless scalars properly represents
the qualitative features of the 4D vacuum energy of the real world,
despite its frequent appearance in the literature.
It was shown in Ref.\cite{Ho:2017joh} that different models of vacuum energy
lead to quite different models of static black holes. 
We study this model as a case study of the various possibilities. 
To say the least,
it provides a concrete toy model of black holes for the sake of discussions,
among others that were proposed in the literature
\cite{metric-models}. 
It points out an interesting new possibility about how vacuum energy
modifies the black-hole geometry in a significant way.

In fact,
it was shown in \cite{Ho:2017joh} that 
a vacuum state with negative energy would 
in general lead to the wormhole-like structure
(a local minimum of $r$) instead of a horizon. 
If the vacuum energy is negative,
the geometry is expected to share some of the qualitative features 
as those described in this paper.
Black holes with back reaction from other models of vacuum energy
will be studies in more detail in the near future.

Based on the static solutions,
we speculate on what to expect 
in a dynamical process of gravitational collapse
for the same model of vacuum energy.
Typically one assumes that
the initial state is the Minkowski vacuum
with zero vacuum energy in the infinite past.
The vacuum energy becomes negative 
and increases in magnitude
during the collapse because of the increment of curvature.
If there is no energy exchange with matters,
the conservation law for the vacuum energy
implies an outgoing (positive) energy flow for energy conservation.
This is a generalization of the notion of Hawking radiation.
Even an arbitrarily slow collapse without the formation of horizon
involves the generalized Hawking radiation.

For simplicity,
we made the assumption that 
the energy-momentum tensors are conserved separately
for the fluid and the vacuum.
As a result,
the Hawking radiation cannot take away the energy of the fluid.
The fluid remains there even when the total energy becomes zero
(the magnitude of the negative vacuum energy equals the fluid energy).
At the end of such a ``complete evaporation'',
the Schwarzschild radius (both $a$ and $a_0$) goes to zero,
or to the Planck scale,
with the liquid persisting in a large space behind the vanishing neck,
resembling Wheeler's bag of gold \cite{Wheeler-Bag-Of-Gold}.
We have shown in this paper that this cannot happen for static configurations 
with realistic density of the fluid. 
However, this can be realized for dynamical configurations. 
This type of solution has already been observed in numerical analysis \cite{Parentani:1994ij}.

In a fundamental theory including quantum gravity such as the string theory,
there are always (direct or indirect) interactions among different fields. 
The quantum fields in vacuum and the collapsing matter are allowed to exchange energy,
and the mass of the fluid decreases during evaporation,
leading to a reduction in size of the interior space under the neck.
The question now is whether the neck shrinks slower than the interior size of the bag,
so that there is no remnant in the end of a complete evaporation. 
For example, if the evaporation of the black hole is sufficiently slow 
and the solution can be approximated by the static solution in this paper at each moment, 
the size of the neck and that of the interior are almost the same and 
the neck cannot shrink to zero with a large bag behind. 
On the other hand, if the evaporation is very fast and 
the fluid inside cannot settle down to almost static configurations, 
the ``bag of gold'' with a large interior and a Planckian-sized neck 
would appear as in some numerical analyses \cite{Parentani:1994ij}. 
In order to see more details on this process, 
time-dependent solutions of the semi-classical Einstein equation should be considered. 
This is left for future studies.

\subsection*{Acknowledgements}

The authors would like to thank 
Emil~Akhmedov,
Yi-Chun~Chin,
Chong-Sun~Chu,
Hsien-chung~Kao,
Samir~Mathur,
Yutaka~Matsuo,
Andrei~Mironov,
Alexei~Morozov,
Inyong~Park,
Yu-Ping~Wang
and
Shu-Jung~Yang
for discussions.
The work is supported in part by
the Ministry of Science and Technology, R.O.C. 
(project no.~104-2112-M-002-003-MY3)
and by National Taiwan University.


\vskip .8cm
\baselineskip 22pt
\begin{thebibliography}{99}
\itemsep 0pt






\bibitem{Gerlach:1976ji} 
  U.~H.~Gerlach,
  ``The Mechanism of Black Body Radiation from an Incipient Black Hole,''
  Phys.\ Rev.\ D {\bf 14}, 1479 (1976).
  doi:10.1103/PhysRevD.14.1479

\bibitem{FuzzBall}
  O.~Lunin and S.~D.~Mathur,
  ``AdS / CFT duality and the black hole information paradox,''
  Nucl.\ Phys.\ B {\bf 623}, 342 (2002)
  [hep-th/0109154].
  O.~Lunin and S.~D.~Mathur,
  ``Statistical interpretation of Bekenstein entropy for systems with a stretched horizon,''
  Phys.\ Rev.\ Lett.\  {\bf 88}, 211303 (2002)
  [hep-th/0202072].
  S.~D.~Mathur,
  ``Resolving the black hole causality paradox,''
  arXiv:1703.03042 [hep-th].

\bibitem{FuzzBall2}
  O.~Lunin, J.~M.~Maldacena and L.~Maoz,
  ``Gravity solutions for the D1-D5 system with angular momentum,''
  hep-th/0212210.
  S.~D.~Mathur,
  ``The Fuzzball proposal for black holes: An Elementary review,''
  Fortsch.\ Phys.\  {\bf 53} (2005) 793
  [hep-th/0502050].
  V.~Jejjala, O.~Madden, S.~F.~Ross and G.~Titchener,
  ``Non-supersymmetric smooth geometries and D1-D5-P bound states,''
  Phys.\ Rev.\ D {\bf 71} (2005) 124030
  [hep-th/0504181].
  V.~Balasubramanian, E.~G.~Gimon and T.~S.~Levi,
  ``Four Dimensional Black Hole Microstates: From D-branes to Spacetime Foam,''
  JHEP {\bf 0801} (2008) 056
  [hep-th/0606118].
  I.~Bena and N.~P.~Warner,
  ``Black holes, black rings and their microstates,''
  Lect.\ Notes Phys.\  {\bf 755} (2008) 1
  [hep-th/0701216].
  K.~Skenderis and M.~Taylor,
  ``The fuzzball proposal for black holes,''
  Phys.\ Rept.\  {\bf 467} (2008) 117
  [arXiv:0804.0552 [hep-th]].
  I.~Bena, S.~Giusto, E.~J.~Martinec, R.~Russo, M.~Shigemori, D.~Turton and N.~P.~Warner,
  ``Smooth horizonless geometries deep inside the black-hole regime,''
  Phys.\ Rev.\ Lett.\  {\bf 117} (2016) no.20,  201601
  [arXiv:1607.03908 [hep-th]].

\bibitem{Barcelo:2007yk} 
  C.~Barcelo, S.~Liberati, S.~Sonego and M.~Visser,
  ``Fate of Gravitational Collapse in Semiclassical Gravity,''
  Phys.\ Rev.\ D {\bf 77}, 044032 (2008)
  [arXiv:0712.1130 [gr-qc]].
 
\bibitem{Vachaspati:2006ki} 
  T.~Vachaspati, D.~Stojkovic and L.~M.~Krauss,
  ``Observation of incipient black holes and the information loss problem,''
  Phys.\ Rev.\ D {\bf 76}, 024005 (2007)
  [gr-qc/0609024].

\bibitem{Krueger:2008nq} 
  T.~Kruger, M.~Neubert and C.~Wetterich,
  ``Cosmon Lumps and Horizonless Black Holes,''
  Phys.\ Lett.\ B {\bf 663}, 21 (2008)
  doi:10.1016/j.physletb.2008.03.051
  [arXiv:0802.4399 [astro-ph]].

\bibitem{Fayos:2011zza} 
  F.~Fayos and R.~Torres,
  ``A quantum improvement to the gravitational collapse of radiating stars,''
  Class.\ Quant.\ Grav.\  {\bf 28}, 105004 (2011).
  doi:10.1088/0264-9381/28/10/105004



\bibitem{Kawai:2013mda} 
  H.~Kawai, Y.~Matsuo and Y.~Yokokura,
  ``A Self-consistent Model of the Black Hole Evaporation,''
  Int.\ J.\ Mod.\ Phys.\ A {\bf 28}, 1350050 (2013)
  [arXiv:1302.4733 [hep-th]].
  
\bibitem{Kawai:2014afa} 
  H.~Kawai and Y.~Yokokura,
  ``Phenomenological Description of the Interior of the Schwarzschild Black Hole,''
  Int.\ J.\ Mod.\ Phys.\ A {\bf 30}, 1550091 (2015)
  doi:10.1142/S0217751X15500918
  [arXiv:1409.5784 [hep-th]].

\bibitem{Ho:2015fja} 
  P.~M.~Ho,
  ``Comment on Self-Consistent Model of Black Hole Formation and Evaporation,''
  JHEP {\bf 1508}, 096 (2015)
  doi:10.1007/JHEP08(2015)096
  [arXiv:1505.02468 [hep-th]].

\bibitem{Kawai:2015uya} 
  H.~Kawai and Y.~Yokokura,
  ``Interior of Black Holes and Information Recovery,''
  Phys.\ Rev.\ D {\bf 93}, no. 4, 044011 (2016)
  doi:10.1103/PhysRevD.93.044011
  [arXiv:1509.08472 [hep-th]].

\bibitem{Ho:2015vga} 
  P.~M.~Ho,
  ``The Absence of Horizon in Black-Hole Formation,''
  Nucl.\ Phys.\ B {\bf 909}, 394 (2016)
  doi:10.1016/j.nuclphysb.2016.05.016
  [arXiv:1510.07157 [hep-th]].

\bibitem{Ho:2016acf} 
  P.~M.~Ho,
  ``Asymptotic Black Holes,''
  arXiv:1609.05775 [hep-th].

\bibitem{Kawai:2017txu} 
  H.~Kawai and Y.~Yokokura,
  ``A Model of Black Hole Evaporation and 4D Weyl Anomaly,''
  arXiv:1701.03455 [hep-th].


\bibitem{Mersini-Houghton}
  L.~Mersini-Houghton,
  ``Backreaction of Hawking Radiation on a Gravitationally Collapsing Star I: Black Holes?,''
  PLB30496 Phys Lett B, 16 September 2014
  [arXiv:1406.1525 [hep-th]].
  L.~Mersini-Houghton and H.~P.~Pfeiffer,
  ``Back-reaction of the Hawking radiation flux on a gravitationally collapsing star II: Fireworks instead of firewalls,''
  arXiv:1409.1837 [hep-th].

\bibitem{Saini:2015dea} 
  A.~Saini and D.~Stojkovic,
  ``Radiation from a collapsing object is manifestly unitary,''
  Phys.\ Rev.\ Lett.\  {\bf 114}, no. 11, 111301 (2015)
  [arXiv:1503.01487 [gr-qc]].

\bibitem{Baccetti}
  V.~Baccetti, R.~B.~Mann and D.~R.~Terno,
  ``Role of evaporation in gravitational collapse,''
  arXiv:1610.07839 [gr-qc].
  V.~Baccetti, R.~B.~Mann and D.~R.~Terno,
  ``Horizon avoidance in spherically-symmetric collapse,''
  arXiv:1703.09369 [gr-qc].
  V.~Baccetti, R.~B.~Mann and D.~R.~Terno,
  ``Do event horizons exist?,''
  arXiv:1706.01180 [gr-qc].


\bibitem{Ho:2017joh} 
  P.~M.~Ho and Y.~Matsuo,
  ``Static Black Holes With Back Reaction From Vacuum Energy,''
  arXiv:1703.08662 [hep-th].


\bibitem{Mathur:2009hf} 
  S.~D.~Mathur,
  ``The Information paradox: A Pedagogical introduction,''
  Class.\ Quant.\ Grav.\  {\bf 26}, 224001 (2009)
  doi:10.1088/0264-9381/26/22/224001
  [arXiv:0909.1038 [hep-th]].
  
\bibitem{Marolf:2017jkr} 
  D.~Marolf,
  ``The Black Hole information problem: past, present, and future,''
  Rept.\ Prog.\ Phys.\  {\bf 80}, no. 9, 092001 (2017)
  doi:10.1088/1361-6633/aa77cc
  [arXiv:1703.02143 [gr-qc]].

\bibitem{Almheiri:2012rt} 
  A.~Almheiri, D.~Marolf, J.~Polchinski and J.~Sully,
  ``Black Holes: Complementarity or Firewalls?,''
  JHEP {\bf 1302}, 062 (2013)
  doi:10.1007/JHEP02(2013)062
  [arXiv:1207.3123 [hep-th]].


\bibitem{Davies:1976ei} 
  P.~C.~W.~Davies, S.~A.~Fulling and W.~G.~Unruh,
  ``Energy-Momentum Tensor Near an Evaporating Black Hole,''
  Phys.\ Rev.\ D {\bf 13}, 2720 (1976).
  doi:10.1103/PhysRevD.13.2720

\bibitem{Christensen:1977jc} 
  S.~M.~Christensen and S.~A.~Fulling,
  ``Trace Anomalies and the Hawking Effect,''
  Phys.\ Rev.\ D {\bf 15}, 2088 (1977).
  doi:10.1103/PhysRevD.15.2088

\bibitem{Parentani:1994ij} 
  R.~Parentani and T.~Piran,
  ``The Internal geometry of an evaporating black hole,''
  Phys.\ Rev.\ Lett.\  {\bf 73}, 2805 (1994)
  doi:10.1103/PhysRevLett.73.2805
  [hep-th/9405007].

\bibitem{Brout:1995rd} 
  R.~Brout, S.~Massar, R.~Parentani and P.~Spindel,
  ``A Primer for black hole quantum physics,''
  Phys.\ Rept.\  {\bf 260}, 329 (1995)
  doi:10.1016/0370-1573(95)00008-5
  [arXiv:0710.4345 [gr-qc]].

\bibitem{Ayal:1997ab}
  S.~Ayal and T.~Piran,
  Phys.\ Rev.\ D {\bf 56} (1997) 4768
  doi:10.1103/PhysRevD.56.4768
  [gr-qc/9704027].

\bibitem{Fabbri:2005nt} 
  A.~Fabbri, S.~Farese, J.~Navarro-Salas, G.~J.~Olmo and H.~Sanchis-Alepuz,
  ``Semiclassical zero-temperature corrections to Schwarzschild spacetime and holography,''
  Phys.\ Rev.\ D {\bf 73} (2006) 104023
  doi:10.1103/PhysRevD.73.104023
  [hep-th/0512167]. 
  A.~Fabbri, S.~Farese, J.~Navarro-Salas, G.~J.~Olmo and H.~Sanchis-Alepuz,
  ``Static quantum corrections to the Schwarzschild spacetime,''
  J.\ Phys.\ Conf.\ Ser.\  {\bf 33}, 457 (2006)
  doi:10.1088/1742-6596/33/1/059
  [hep-th/0512179].


\bibitem{Solodukhin:2004rv}
  S.~N.~Solodukhin,
  ``Can black hole relax unitarily?,''
  hep-th/0406130.
  S.~N.~Solodukhin,
  ``Restoring unitarity in BTZ black hole,''
  Phys.\ Rev.\ D {\bf 71} (2005) 064006
  [hep-th/0501053].
  T.~Damour and S.~N.~Solodukhin,
  ``Wormholes as black hole foils,''
  Phys.\ Rev.\ D {\bf 76} (2007) 024016
  [arXiv:0704.2667 [gr-qc]].


\bibitem{StochasticGravity}
  S.~Sinha, A.~Raval and B.~L.~Hu,
  ``Black hole fluctuations and back reaction in stochastic gravity,''
  Found.\ Phys.\  {\bf 33}, 37 (2003)
  doi:10.1023/A:1022815724856
  [gr-qc/0210013].
  B.~L.~Hu and E.~Verdaguer,
  ``Stochastic gravity: Theory and applications,''
  Living Rev.\ Rel.\  {\bf 7}, 3 (2004)
  doi:10.12942/lrr-2004-3
  [gr-qc/0307032].


\bibitem{Davies:1976hi} 
  P.~C.~W.~Davies and S.~A.~Fulling,
  ``Radiation from a moving mirror in two-dimensional space-time conformal anomaly,''
  Proc.\ Roy.\ Soc.\ Lond.\ A {\bf 348}, 393 (1976).


\bibitem{Boulware}
  D.~G.~Boulware,
  ``Quantum Field Theory in Schwarzschild and Rindler Spaces,''
  Phys.\ Rev.\ D {\bf 11}, 1404 (1975).
  doi:10.1103/PhysRevD.11.1404
  D.~G.~Boulware,
  ``Hawking Radiation and Thin Shells,''
  Phys.\ Rev.\ D {\bf 13}, 2169 (1976).
  doi:10.1103/PhysRevD.13.2169


\bibitem{Wheeler-Bag-Of-Gold}
J. A. Wheeler, in Relativity, Groups and Topology,
edited by B. DeWitt and C. DeWitt,
p.408 -- 31, Gordon and Breach (1974).


\bibitem{Carballo-Rubio:2017tlh}
  R.~Carballo-Rubio,
  ``Stellar equilibrium in semiclassical gravity,''
  arXiv:1706.05379 [gr-qc].


\bibitem{Buchdahl:1959zz} 
  H.~A.~Buchdahl,
  ``General Relativistic Fluid Spheres,''
  Phys.\ Rev.\  {\bf 116}, 1027 (1959).
  doi:10.1103/PhysRev.116.1027


\bibitem{Park:2017dib} 
  I.~Y.~Park,
  ``Quantum-corrected geometry of horizon vicinity,''
  arXiv:1704.04685 [hep-th].


\bibitem{metric-models}
  S.~A.~Hayward,
  ``Formation and evaporation of regular black holes,''
  Phys.\ Rev.\ Lett.\  {\bf 96}, 031103 (2006)
  doi:10.1103/PhysRevLett.96.031103
  [gr-qc/0506126].
  A.~Bonanno and M.~Reuter,
  ``Spacetime structure of an evaporating black hole in quantum gravity,''
  Phys.\ Rev.\ D {\bf 73}, 083005 (2006)
  doi:10.1103/PhysRevD.73.083005
  [hep-th/0602159].
  V.~P.~Frolov,
  ``Information loss problem and a 'black hole` model with a closed apparent horizon,''
  JHEP {\bf 1405}, 049 (2014)
  doi:10.1007/JHEP05(2014)049
  [arXiv:1402.5446 [hep-th]].
  J.~M.~Bardeen,
  ``Black hole evaporation without an event horizon,''
  arXiv:1406.4098 [gr-qc].



\end{thebibliography}
\end{document}